\documentclass[12pt,preprint]{aastex}
\usepackage{graphicx}

\def\4258{NGC~4258}
\def\XMM{XMM-{\it Newton}}
\def\Chandra{{\it Chandra}}
\def\ASCA{{\it ASCA}}
\def\SAX{{\it Beppo}-SAX}
\newcommand{\etal}{{~et al.}~}

\slugcomment{Accepted by The Astrophysical Journal}

\shorttitle{XMM and Chandra Observations of NGC~4258}
\shortauthors{Fruscione\etal}

\begin{document}

\title{X-ray Luminosity and Absorption Column Fluctuations 
in the  H$_2$O Maser Galaxy NGC~4258 from Weeks to Years }

\author{Antonella Fruscione\altaffilmark{1}, 
        Lincoln J. Greenhill\altaffilmark{1,2},
        Alexei V. Filippenko\altaffilmark{3},
        James M. Moran\altaffilmark{1},
        James R. Herrnstein\altaffilmark{4}, 
        Elizabeth Galle\altaffilmark{1}}
\altaffiltext{1}{Harvard-Smithsonian Center for Astrophysics,
60 Garden Street, Cambridge, MA 02138; afruscione@cfa.harvard.edu;
greenhill@cfa.harvard.edu; moran@cfa.harvard.edu; egalle@cfa.harvard.edu}

\altaffiltext{2}{Visiting physicist, Kavli Institute for Particle Astrophysics and Cosmology,
Stanford Linear Accelerator Center, M.S. 29, 2575 Sand Hill Rd, Menlo
Park, CA 94025}
 
\altaffiltext{3}{Department of Astronomy, University of California,
Berkeley, CA 94720-3411; alex@astro.berkeley.edu}

\altaffiltext{4}{Renaissance Technologies, Inc., 600 Rt. 25A,
E. Setauket, NY 11733; jrh@rentec.com} 

\begin{abstract} 
We report monitoring of the 0.3--10~keV spectrum of \4258~ with the the
\XMM\ Observatory at five epochs over 1.5 years.  We also report
reprocessing of an overlapping four-epoch series of archival
\Chandra\ observations (0.5--10~keV). 
By including earlier \ASCA\ and \SAX\ observations, we present a new,
nine-year time-series of models fit to the X-ray spectrum of \4258.
We model the \Chandra\ and \XMM\ data self-consistently with partially
absorbed, hard power-law, soft thermal plasma, and soft power-law
components.  Over the nine years, the photoelectric absorbing column
($\sim 10^{23}$~cm$^{-2}$) did not vary detectably, except for a $\sim
40\%$ drop between two \ASCA\ epochs separated by 3 years (in 1993 and
1996) and a $\sim 60\%$ rise between two \XMM\ epochs separated by
just 5 months (in 2001 and 2002).  In contrast, factor of 2--3 changes
are seen in absorbed flux on the timescale of years.  These are
uncorrelated with changes in absorbing column and indicative of
central engine variability.  The most rapid change in luminosity
(5--10~keV) that we detect (with \XMM\ and \Chandra) is $\sim$30\% 
over 19 days.  
The warped disk, a known source of H$_2$O maser emission in
\4258, is believed to cross the line of sight to the central engine.
We propose that the variations in absorbing column 
arise from inhomogeneities sweeping across the line of sight in the rotating disk 
at the radius where the disk crosses the
line of sight.
We estimate that the inhomogeneities are $\sim 10^{15}$~cm in size at the 
crossing radius of $0.29$~pc,
slightly smaller than the expected scale height of the disk. This result thus provides strong  
evidence that the warped accretion disk is the absorber in this
(and possibly other) active galactic nuclei (AGNs). This is the first direct confirmation 
that obscuration in type-2 AGN may, in some cases, 
arise in thin, warped accretion disks, rather than in geometrically thick tori.
Some previous studies report detection of weak Fe K$\alpha$ emission in
\4258.  We do not detect this line emission in any of our \XMM\ spectra
with a 90\% upper limit to the equivalent width of $\sim 49$ eV for
one observation.  Weak, time-variable Fe line absorption has also been
reported for a previous
\Chandra~study.  We do not observe evidence of absorption lines in the
\XMM\ or reprocessed \Chandra\ data. The absence of Fe line emission
is consistent with the disk being optically thin to hard photons as
well as subtending a small solid angle as seen from the central
engine because of the known shallowness of the warp. 
\end{abstract}


\section{Introduction}
\label{s:intro}

Radio interferometric imaging studies of H$_2$O maser emission have
directly demonstrated that, in at least a few cases, type-2 active galactic
nuclei (AGNs) contain thin, nearly edge-on, warped accretion
disks, with large columns of molecular gas inside $\sim 1$~pc. Examples include
NGC~4258 (Miyoshi et al. 1995), NGC~1068
\citep{gg97}, and the Circinus galaxy \citep{greenhill03}.  
In these systems, the maser emission is likely stimulated by
irradiation and heating of relatively cold gas by hard X-rays
\citep[e.g.,][]{nm95}.  Because the structure of the maser emission
can be readily resolved in velocity and angle, it has been possible to
use it to map dense ($n_{\rm H_2}\approx 10^{9\pm 1}$ cm$^{-3}$), warm
($T\approx 300-1000$\,K) accretion-disk gas
\citep{elitzur92}.  

The H$_2$O source in the low luminosity ($L/L_{Eddington}\approx
10^{-4}$) AGN of NGC\,4258 (M106) is remarkable because the disk is
extremely well described by a relatively simple dynamical model
\citep{miyoshi95}.  The maser emission has been mapped from radii of
0.14\,pc to 0.28\,pc around a $3.9\times10^7$\,M$_\odot$ central mass.
Maser positions, line-of-sight velocities and accelerations, and
proper motions are all measurable, and they fit a simple geometrically
thin, slightly warped, Keplerian disk model to high accuracy
\citep{herrnstein99}.  The knowledge of these quantities presents
strong constraints for hard X-ray studies aimed at understanding
processes related to the central engine: emission
mechanisms and surrounding disk structure.

The structure and energetics of type-2 AGNs, in fact, are still not well understood
mainly because X-ray spectra are complex superpositions of components
due to several processes (compact and extended emission, absorption,
scattering, and reflection with different ionization states) and most
astronomical wavebands currently cannot resolve structures $< 1$~pc
in size.  Many AGN models postulate that all AGN are intrinsically
the same and orientation plus obscuration effects regulate their
classification \citep[e.g.,][]{antonucci85, urry95, elvis00}.  However, the actual
nature of the obscuring material is still very much under debate, and several 
locations and structures have been proposed. These include a geometrically thick 
structure, galactic dust \citep{malkan98}, a geometrically thin absorber confined 
to radii comparable 
to the broad-line regions \citep[e.g.,][]{madejski00,
risaliti02}, or an outflow (Elvis \etal\ 2000).

In the case of \4258\ the explanation for the unusually low luminosity
is also an unsettled question: is it the result of a low accretion rate
or radiatively inefficient accretion? \cite{nm95} suggested the
former. Other authors \citep[e.g.,][]{lasota96,gammie99} have
modeled the global spectral energy distribution of \4258\ with
advection-dominated, radiatively inefficient accretion flows (ADAF).
\cite{yuan02} have suggested an alternative model in which outflow
from the ADAF powers the jet in \4258, the emission from which
dominates the continuum from radio to X-ray wavebands. In each case the
hard X-ray continuum is expected to arise from a different location
(a corona above the disk, hot plasma centered on the central engine,
or the base of a jet).

While early soft ($<2.5$ keV) X-ray observations detected emission
from \4258\ (Fabbiano, Kim, \& Trinchieri 1992; Pietsch \etal\ 1994),
only \ASCA\ X-ray observations in the hard band (2--10~keV) allowed
\citet{makishima94} to model for the first time the spectrum of the
central X-ray source of \4258\ as an absorbed power law, weak Fe
K$\alpha$ line, and optically thin soft thermal component.
\citet{reynolds00} report on later \ASCA\ observations and temporal
variability on time scales of years in the absorbing column (30\%) and
in the 5--10~keV flux (a factor of two). Observations with the
$Beppo$-SAX PDS detected an extension of the power-law component to
70~keV with no significant reflection component:
\4258\ is one of the only five Seyfert 2 galaxies, among a total of 31,
for which model fits to the \SAX\ data are not improved by the
addition of a reflection component (Risaliti 2002);  \SAX\ 
MECS observations detected a
factor of two variations in count rate (3--10~keV) on time scales
$<100$~ks \citep{fiore01}, though
\citet{reynolds00} did not report variability over 200~ks for a later epoch. 

Observations of NGC~4258 with \Chandra\ and \XMM\ have been the
first to separate the central compact emission component from the bulk
of the extended soft emission that originates in well-known
nonstellar spiral arms on kpc scales \citep[e.g.,][]{wyc01}. The
first \Chandra\ and \XMM\ spectra are in agreement with \ASCA\ and
\SAX\ spectra, except that Fe K$\alpha$ emission is not detected,
which has raised doubts about the original detections.  Nonetheless,
the line emission is presumed to be closely associated with the
central engine, and non-detection may be indicative of variability in
line equivalent width of a factor of two on time scales of 1--2
years.  Time variability and narrow line-width 
would argue in favor of an association between the line emission and
accretion-disk material \citep{reynolds00} at radii between
$\sim3\times10^{-4}$ and $\sim1$~pc, either by reflection of light
from a variable central source or by variation in disk structure
\citep{pietsch02}.  Variability on shorter time scales ($\sim 10$~ks) has
been claimed by \citet{young04} in the possible detection of weak absorption
features at 6.4 and 6.9~keV, which presumably would be caused by disk
or outflow material at radii $\ll 1$~pc.

We report in this paper the monitoring of NGC~4258 at five epochs with
\XMM\ (including the epoch
reported by Pietsch \& Read 2002) and four epochs with \Chandra\
\citep{young04} that are publically available.  In total, the \XMM\
and \Chandra\ monitoring covers 2.2 years, 2000.2 to 2002.4.  The
primary goal of the program was characterization of time variability
in the absorbing column that might be related to substructure of the
accretion disk traced by maser emission.  We present a time series of
spectral models for NGC~4258, where we have minimized sources of
systematic uncertainty by reprocessing and reanalyzing all existing~
\Chandra\ and \XMM\ datasets using a consistent set of models and
techniques, as well as the most recent calibration databases.  We
describe the \XMM\ and \Chandra\ datasets, their calibration, and the
spectral fitting in Section~2. We discuss our findings in
Section~3. UT dates are used through this paper.

\section{Observations and Analysis}
\label{s:obs}
\subsection{\XMM\ Observations}
NGC~4258 was observed four times from 2001 May to 2002 May as part of
an \XMM\ GO program. Table 1 lists the observation log for the EPIC
instrument including the useful exposure times before and after
filtering for high background times following the procedure outlined
below. For completeness and consistency in analysis, we have included
the first \XMM\ observation of \4258, obtained in 2000 December
\citep{pietsch02}.

\begin{figure}
\epsscale{1.0}
\plotone{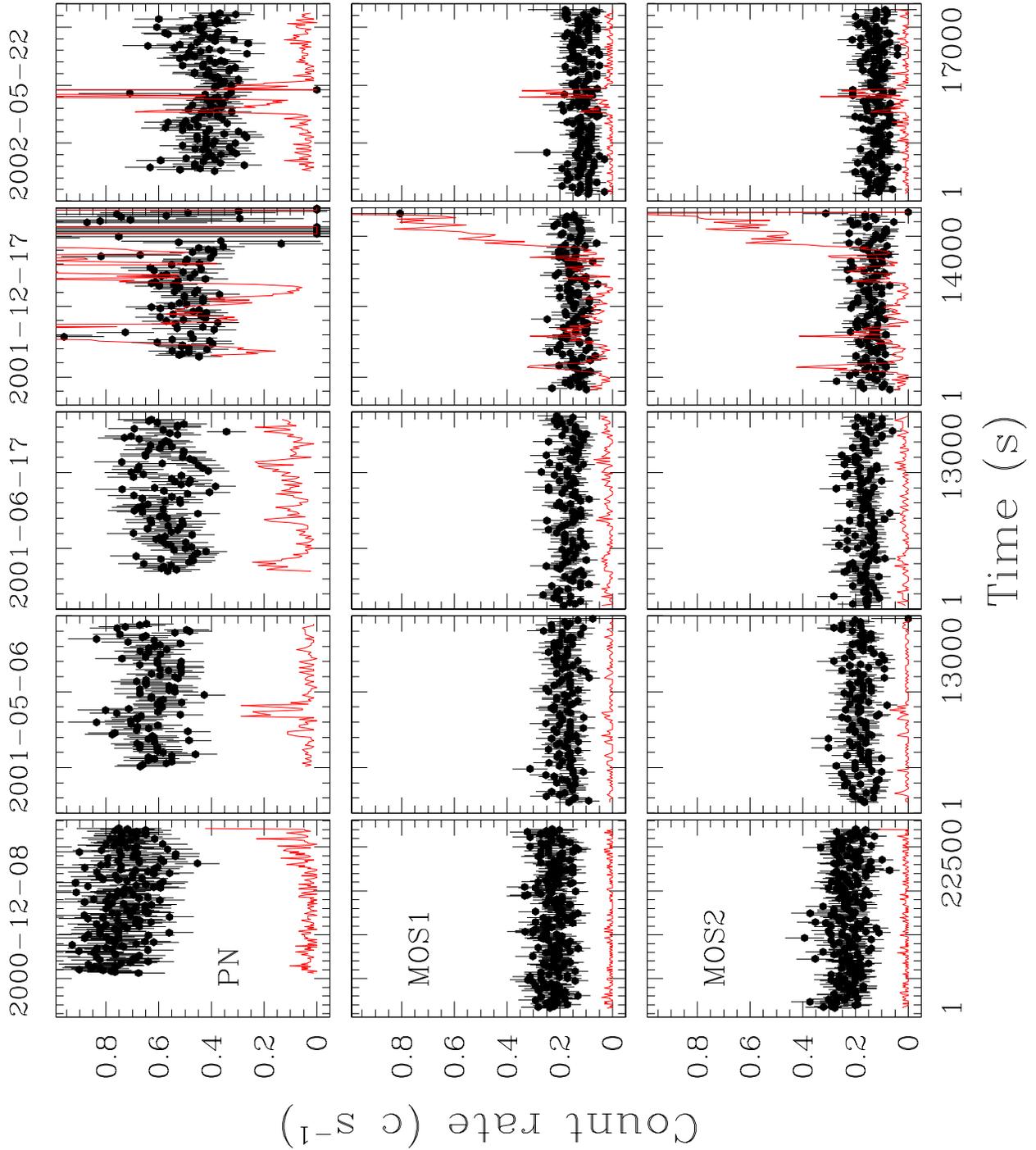}
\caption{\XMM\ EPIC background-subtracted light curves binned in 100 s bins for all five observation epochs. 
PN, MOS1, and MOS2 data are shown in the three rows. The continuous lines represent the level of the background. 
The light curves were extracted for the nuclear region of \4258\ (15\arcsec\ extraction radius) over the entire instrumental energy range (nominally 0.2--15 keV). Note that during epoch 4 (2001-12-17) the observation was affected by strong background flares and a generally very high background level. Also note that the time spacing is {\it not} uniform between the different epochs.
}
\label{f:lightcurve0}
\end{figure}

All EPIC PN and MOS observations were taken in Prime Full Window mode
with a Medium Filter.  The entire data analysis was accomplished with
a combination of tasks in the \XMM\ and \Chandra\ data analysis
software systems (Science Analysis Software - SAS - versions
20030110\_1802/20040318\_1831 and \Chandra\ Interactive Analysis of
Observations - CIAO - versions 3.0.1/3.0.2).  We generated calibrated
EPIC event files using the {\it epproc} and {\it emproc} routines and
filtered the final event list to include only ``good'' events,  
with patterns 0--4 (singles and doubles) for the PN camera and 0--12 for the 
MOS cameras.
We also rejected bad times due to high background flares
(Figure~\ref{f:lightcurve0}) through a comparison of the source and
background light curves for each observation. Ratios of source and
background count rates were calculated for each 100~s time-bin, and we
eliminated all time intervals with source-to-background count ratios
lower than 5.  Useful integration times were reduced as shown in Table
1.  The observation most affected by overall high background levels as
well as strong flares was on 2001 Dec. 17 (Figure~\ref{f:lightcurve0}).
We compared the expected and observed distributions of event grades
calculated with the SAS routine {\it epatplot} to quantify the effects
of event pile-up, which were not detectable.\footnote{When two (or more)
photons are detected in a single pixel within one read-out period,
they are perceived to be a single, ``piled-up'' event, and the count
rate is depressed.  The resulting distribution of grades and the
energy spectrum are distorted since the apparent energy is
approximately the sum of the photon energies.}

Spectra were extracted using the SAS routine {\it especget} excluding
events near the edge of the CCDs and near bad pixels which may have
incorrect energies.  We chose a circular source extraction region of
300 pixel radius (or 15\arcsec) which covers about 80\% of the PSF and
a background extraction region of 1200 pixel radius (60\arcsec) lying
$\sim6$\arcmin\ from the source on the same chip as the source region.
We binned the spectral pulse-invariant (PI) channels by a factor of 10
in all spectra to obtain uniform binning while increasing the
statistics in each individual bin (Figures~2--6).

The most recent response matrices (version of 2003-01-29) for each
instrument observing mode and observation date were obtained from the
\XMM\ calibration archive and used in the spectral fitting.  The same
source and background extraction regions were used for deriving the
light curves (Figure 1) for the PN and MOS cameras. All light curves
were extracted over the entire energy range (nominally 0.2--15 keV)
taking into account an average dead-time correction calculated from
the {\it livetime} and {\it ontime} keywords recorded in the data.
Average count rates for all observations are listed in Table 1.

We note in passing that we also checked the \XMM\ RGS spectra which, however, 
provided no additional constraints to the spectral modeling because 
of the extremely poor signal-to-noise ratio in the covered 
waveband ($<2.5$ keV).

\subsection{\Chandra\ Observations}

The \Chandra\ ACIS-S camera observed \4258\ four times from 2000 March
to 2001 May (Table 2) with useful exposure times ranging from
$\sim1.6$ to $\sim20$ ks, as described by
\citet{pietsch02} and \citet{young04}. 
The 2000 April 17 (13~ks) and 2001 May 29 (20~ks)
observations were  optimized to study the faint ``anomalous arms'' of \4258.  
As a result, both  suffer from significant photon pile-up in the nuclear region, 
of order 20--30\%. Detailed modeling was needed
to account for this effect when fitting spectra (see $\S2.3.1$).

We re-analyzed all \Chandra\ data to ensure that the most recent calibrations
were used and that the analysis techniques were consistent with those used for the
\XMM\ data. We reprocessed the data using the
\Chandra\ calibration database (CALDB) versions 2.23 and 2.26 and CIAO software
version 3.0.1 and 3.0.2.  
In the data preparation phase we checked for background
flares (none present in any of the data sets) and defined extraction
regions: a 2\arcsec\ radius source region on the nucleus and a 10\arcsec\
radius background region about 2\arcmin\ from the nucleus.
Responses were derived from the latest calibration data available in
CALDB 2.26 and included the effects due to a reduction of the low-energy 
quantum efficiency of the ACIS instrument\footnote{More details  
can be found at http://cxc.harvard.edu/cal/Acis/Cal\_prods/qeDeg/.}.  
As in the \XMM\
analysis, extracted spectra were binned by a factor of 10 in spectral
PI channels (Figures ~7, 8, 11, 12). In some cases we verified the results by also binning to
$\geq15$ counts per bin for comparison.

\subsection{Spectral Analysis}
We used version 3.0.2 of {\it Sherpa} (Freeman, Doe, \& Siemiginowska
2001) to fit models to our spectra iteratively, widening the energy
interval being fit and adding model components as required to improve
the significance of the spectral fit (Figures 2--8, 11, 12) .

For the \XMM\ data we started by fitting the 2--10 keV energy range
with a simple power-law model, absorbed by both intrinsic absorption
and Galactic absorption ($N_{H}$), the latter fixed at the value
$1.19\times10^{20}$ cm$^{-2}$ (Murphy et al. 1996).  We verified that
the background is negligible for all observations, so we neither
subtracted it nor modeled it during the spectral fitting.  We then
extended the energy range to 0.3~keV, adding a thermal plasma component
(the ``mekal'' model: Mewe, Gronenschild,
\& vad den Oord 1985; Mewe, Lemen, \& van den Oord 1986; Kaastra 1992; Liedahl, Osterheld, \&
Goldstein 1995) with elemental abundances as a fixed fraction of 
the cosmic values (Anders \& Grevess 1989), and a soft power-law
component, both absorbed by the Galactic $N_{H}$.  The best-fit values for
all the free parameters in the models are listed in Table 2, together
with model-derived fluxes and luminosities in various energy ranges,
assuming a distance of 7.2
Mpc (Herrnstein et al. 1999) so 1\arcsec\ = 35 pc.

In order to assess the quality of the fit we used the $\chi^2$
statistic proposed by Kearns, Primini,
\& Alexander (1995) and implemented in {\it
Sherpa} (so-called ``chi primini'' statistic) which is appropriate for 
non-background-subtracted data.
This statistic can be used even when the number of counts in each bin is small ($<5$) and it 
is not biased (see, e.g., Freeman 2001).

Spectral fits obtained with a bremsstrahlung component instead of a
soft power-law component work equally well at soft energies and
neither has a significant influence on the best-fit parameters for the
direct AGN component (both in term of slope and absorption). These
soft components may reflect some integrated Galactic contribution,
such as from the X-ray binary population.  The thermal plasma component is
needed to account for the strong iron L complex detected at $\sim1$ keV.

For the \Chandra\ observations the background was also negligible, so
we neither subtracted it nor modeled it. However, in any case, we
note that background subtraction is not recommended (Davis 2001) for
data to which pile-up models will be applied.  We again started the analysis by
fitting a simple absorbed power-law model in the 2--10 keV energy
range. We then added two model components to extend the fit to 0.5
keV.  Finally, for the two longest \Chandra\ observations, we added a
pile-up model to the fit (see below).  The ``chi primini'' statistic
was used for the two shortest \Chandra\ observations, while the ``chi
dvar'' statistic (a $\chi^2$ statistic with variance computed from
the observed counts data) was preferred for the two longest piled-up
\Chandra\ observations for computer efficiency reasons.

Inspection of the residuals to the best fits to \XMM\ and
\Chandra\ data did not reveal the presence of statistically significant
emission or absorption lines in any of the spectra. To calculate upper
limits to Fe K$\alpha$ emission line equivalent widths, we added a
Gaussian component to the 2--10 keV power-law model (Table~2), fixing
the line energy ($E_{line}$) at 6.4~keV and obtaining upper limits to
the equivalent widths ($W_{line}$) using the estimated upper bound of
the uncertainty in line normalization (90\% confidence).  In one case
(the 2001 June 17 \XMM\ observation) a line can be fit, if $E_{line}$
is allowed to vary (Table 2); however, the fit is only improved at $<2 \sigma$
level by the
addition of the Gaussian model component (reduced $\chi^2$=1.05 for
159 degrees of freedom without a line, reduced $\chi^2$=1.02 for 156
degrees of freedom with a line), which we therefore do not classify as a detection.

\subsubsection{Pile-up Modeling of \Chandra\ Data}

As noted earlier, two of the \Chandra\ observations (on 2000 April 17
and 2001 May 28) were strongly affected by photon pile-up. We
accounted for this at the spectral fitting level by including the
model by Davis (2001), as implemented in {\it Sherpa}, in our overall
spectral fit.  Following the instructions for using the model
for \Chandra\ data we

(i) re-included all events that the automated
pipeline classifies as ``afterglow'' events before performing the
spectral extraction,

(ii) verified the constancy of the source during
the observation,

(iii) extracted the spectra from a 2\arcsec\ radius
source region around the nucleus, and 

(iv) did not subtract the
background in the fitting process.  

The Davis (2001) pile-up model is described by various parameters, two of which
are generally the most crucial: $\alpha$ (which represents the
probability of getting a ``good'' event grade --- essentially a
non-background event --- when two photons pile together), and $f$ (the
fraction of flux falling into the pile-up region).  To investigate
systematic errors that may result from the details of how pile-up is
modeled, we processed each pile-up epoch in three ways: with $\alpha$
and $f$ free to vary, $\alpha$ free and $f$ fixed, and $\alpha$ and
$f$ both fixed.  The results of the fits are reported in Table 3.
For both epochs the three pile-up corrected models are statistically
consistent within the 90\% errors and the global pile-up fraction
(i.e. the estimated percentage of observed counts containing two or
more photons piled together) is not significantly affected by the
choice of $\alpha$ and $f$ parameters.  However, we note a systematic
bias in parameters when either $\alpha$ or $f$ are held fixed (e.g., we
obtain a steeper photon index, $\Gamma_{pl}$). 
We also note better agreement with the
non-piled-up \Chandra\ epochs, and lower reduced $\chi^2$,  when $\alpha$ and $f$
are allowed to vary during the fit.
For this reason we adopt ``unfrozen'' models and from a comparison of
parameters listed in Table~2 suggest that the systematic uncertainties
in $L_{2-10}$, $\Gamma_{pl}$, and $N_{H}$ are on the order of 20\%, 20\%, and
10\%.  (Hard absorbed flux is not significantly affected.)
Differences between our best-fit parameters and those of \citet{young04} are 
probably in part related to the treatment of pile-up\footnote{See also Nowak 2003 at
http://space.mit.edu/CXC/analysis/PILECOMP/ for a detailed discussion of modeling piled-up
spectra with high-energy tails.}.

\subsection{Comparison of Results Obtained for Different Missions}

In Table~2 we report a cumulative record of spectral parameters from
the literature and obtained by us in processing new and archival
\Chandra\ and \XMM\ datasets.  Care in comparison of results obtained
with different instruments is warranted because of different angular
and energy resolutions, energy coverage, and cross-calibration effects (see,
e.g., Marshall \etal\ 2001 on \Chandra-\XMM\ cross-calibration).  
For example,
analysis of \SAX\ data (extraction radius 3\arcmin\ in the 2--10 keV band)
has delivered the steepest $\Gamma_{pl}$, and
\SAX\ was effectively the only mission sensitive to emission above
$\sim 10$~keV.  Analyses of \Chandra\ data consistently provide the
shallow estimates of $\Gamma_{pl}$ and low estimates of $N_H$ (for
epochs free of pile-up and epochs where pile-up parameters are free to
vary; see, e.g., Figure~\ref{f:confidence}).  This may in part be a
consequence of angular resolution --- in other words, a small extraction radius
(2\arcsec\ or $\sim 70$~pc) better isolates the central engine from
surrounding soft nuclear emission. (\XMM\ supports the next smallest
extraction radius, 15\arcsec\ or $\sim 520$~pc.)  On the other hand,
hard and soft components can be disentangled in analysis of somewhat
lower angular resolution data (e.g., \XMM) if the source model is
accurate.  In that case, the limited effective collecting area of
\Chandra\ above $\sim 6$~keV may be a more significant factor, as it
reduces constraints on the hard portion of the spectrum and increases
correlations between fit parameters (such as $N_H$ and $\Gamma_{pl}$).
In contrast, \XMM\ count rates are higher, provide better constraint
on the hard portion of the spectrum, and probably provide more robust
parameter estimates (for this galaxy).

\section{Discussion}
\label{s:discuss}

We have added four new epochs to the monitoring record of hard X-ray
emission from \4258\, which now extends to 9 years.  The five
\XMM\ observations by themselves provide a
much needed time-series of 0.3--10~keV spectra with high count rate and
spectral resolutions that are unaffected by serious instrumental
systematics, such as pile-up.  The time separations between \XMM\
epochs span 0.12 to 1.5 years. Taken together, the time series of
\Chandra\ and \XMM\ epochs spans 2.2 years with a minimum separation 
of 0.05 years (3 weeks).

The source models fit to the \XMM\ data are on average consistent with
the range of models fit to data from other satellites.  For the five
\XMM\ epochs, we obtain an absorbing column
$N_H = (8.6$--$13.2) \times 10^{22}$~cm$^{-2}$, absorption-corrected
luminosity $L_X = (5.4$--$9.1) \times 10^{40}$~ergs s$^{-1}$ (2--10~keV) or
$L_X = (2.9$--$4.4) \times 10^{40}$~ergs s$^{-1}$ (5--10~keV), photon index
$\Gamma_{pl}=1.5$--1.7, and thermal plasma temperature $kT = 0.5$--0.6~keV
(Table~2). The characteristics of the soft emission component do not
in general display substantial range, and the \XMM\ results are
statistically consistent with estimates reported elsewhere.

\subsection{Continuum Variability}

Count rates and estimates of absorbed flux ($f_X$) obtained just with
\XMM~(2--10~keV and 5--10~keV) demonstrate intensity variations of 10--30\%
on time scales as short as six months (Figures~\ref{f:lightcurve0},
\ref{f:trends}; Table~2).   However, comparing \XMM\ and \Chandra\ data that
are unaffected by pile-up, we observe a $\sim 30\% $ decrease in
hard (5--10~keV) absorbed flux between epochs separated by 19~days (2001
May 29 and  2001 Jun 17). Count rates estimated by \citet{young04} also
appear to indicate variation between two epochs separated by $\sim 1$~month
(2000 March 8 and 2000 April 17).  However, interpretation of this
comparison requires caution because of systematic uncertainty associated
with pile-up modeling applied to the later epoch (see $\S2.3.1$).

Evidence of more rapid variability has been reported, but the
contributions of systematics can be difficult to assess, such as pile-up 
(in \Chandra\ data),
background variability, or the presence of other variable sources in the
extraction region. For example, \citet{fiore01} report \SAX\ light-curve
variations of 10--20\% on  time scales of $\sim 1$ hour and 100\% on time
scales on the order of half a day,  but without background subtraction and for
an extraction region of 3\arcmin\ (which includes, for example, the soft
companion discovered by \Chandra\ $\sim3$\arcsec\ to the southwest of the nucleus;
Pietsch \& Read 2002). In contrast, on time scales of $\sim 3$~hours, we
observe with \XMM\ no variability appearing in background-subtracted light
curves for both the PN and MOS instruments (Figure~\ref{f:lightcurve0}).
Similarly, \citet{reynolds00} observe no
variability with \ASCA~on scales up to
$\sim 2$~days.

We also detect variability in unabsorbed luminosity ($L_X$) with
\XMM\ (Figure~\ref{f:trends}).  In the $\sim 1$~year prior to the last epoch, 
$L_X$ (2--10~keV) dropped $\sim 20\%$ to $\sim 5.4\times10^{40}$~erg
s$^{-1}$, possibly as part of a long-term decline by approximately a
factor of two from the peak detected by \Chandra\ on 2000 March 8. This large
variation is comparable in magnitude to what has been
seen for \ASCA,
\SAX, and early \Chandra\ epochs  (Figure~\ref{f:trends}), when
$L_X$(2--10~keV) varied by factors of $\sim2$ \citep[see
also][]{reynolds00}.  We exclude the possibility that cross-correlation in
the modeling causes these variations, referring to Figure~\ref{f:trends} and
the absence of any correlation between estimates of $L_X$ and $N_H$. (We
recall that $N_H$ and $\Gamma$ appear to be somewhat correlated among models
for \Chandra~data (see $\S2.3.2$), but emphasize that over a wide range,
variations in $L_X$ do not appear to be correlated with changes in
either $N_H$ or $\Gamma$.) 

Although there is no one-to-one correspondence, we note that two of the
lowest estimates of $L_X$ are associated with the two largest estimates of
$N_H$. Within the $\sim 5$~months prior to the last \XMM\ epoch, $N_H$
increased $\sim 50\%$, returning to a high
level [($1.3 \pm 0.1) \times10^{23}$~cm$^{-2}$] not reported since the first
\ASCA\ observations in 1993.  In both cases, $L_X$(5--10~keV) is
$<4\times10^{40}$~ergs~s$^{-1}$.  We speculate that reductions in $L_X$
might affect the ionization state of the absorber, resulting in increased
columns in some cases, though we acknowledge that \4258\ is a weak source,
and it is difficult to establish a connection. If true, then a change in
ionization would suggest that the absorber is not entirely neutral.

\subsection{Absorbing Gas}

\4258\ is an unusual target to model because the central mass, 
geometry, and dynamics of surrounding dense material are accurately
known from modeling of radio interferometry data of the maser emission.  The best-fit
warped disk model, which is derived from observed maser positions and Doppler
velocities, is highly inclined. In particular, the warp of the disk
brings it across the line of sight to the dynamical center at a radius of 0.29 pc, where 
its angle to the line of sight is about $16^\circ$, just beyond the
outer limit of the known maser emission \citep{herrnstein04}, and 
thus placing high-density gas ($\sim 10^{9\pm1}$~cm$^{-3}$;
Elitzur 1992) between us and the central engine. 

To interpret the time variability in the absorption column density, we
assume that inhomogeneities in the  disk move
across our line of sight as the disk rotates and that these are
responsible for observed changes in absorption. The discussion that follows
depends rather critically on the view that the maser emission is
a good tracer of the disk --- i.e., that its emission defines the 
mid-plane of the disk, and that the disk changes from molecular
to atomic beyond a radius of about 0.28 pc, as described by 
\citet{nm95}. For these assumptions, the mass accretion rate
is about $10^{-4} \alpha^{-1}$ solar masses per year.

We estimate the physical size of the inhomogeneities in the disk from the
variability time scale of the absorption column density.  The rotation
speed of the disk at a radius of 0.29 pc is about $v_{rot} = $~760 km~s$^{-1}$ for
a central mass of 3.9 $\times 10^{7}$~M$_\odot$. Hence the length scale ($r_c$) 
associated with the observed time-variability scale ($\tau$) of 5 months 
is $r_c = v_{rot} \tau$,
or about $1.0 \times 10^{15}$ cm.  For the standard thin accretion disk
(e.g., Frank, King, \& Raine 1992), the scale height (h) obeys the
relation $ h = c_s r^{3/2}(GM)^{-1/2}$. If the gas is atomic at
the line-of-sight crossing radius, then the temperature is expected to be about
8000~K and $c_s = 7$ km~s$^{-1}$, according to the model of \citet{nm95}. Hence,
h = $8 \times 10^{15}$ cm and $r_c \approx h/8$. If the gas at the
crossing radius were molecular, and at the preferred maser temperature
of about 800~K, the sound speed would be about 2.3 km~s$^{-1}$, and the
scale height would then be about $1.1 \times 10^{15}$ cm, close to $r_c$.
However, if the gas were atomic, then the mass accretion rate, which 
scales as $ c_{s}^{2}$,  would
have to be about $10^{-5} \alpha^{-1}$ M$_{\odot}~ {\rm yr}^{-1}$. This would move the 
atomic transition radius into the maser region, and also make the 
density at the inner radius of maser emission unacceptably low. 
We note that the time scale of bright maser flares is about 2 months
\citep{herrnstein05}. At the radius of the flare maser, 0.17 pc,
the rotation speed is about 970 km~s$^{-1}$, the scale for
density fluctuations is $5 \times 10^{14}$ cm, and the scale height is
$1.2 \times 10^{14}$ ~cm. Hence, here also the turbulence scale length 
is somewhat smaller than the local scale height.

Advective accretion models have been proposed for
\4258\ \citep[e.g.,][and references therein]{gammie99}  which require much larger accretion
rates and much thicker disks, and for which the masers are
relegated to a thin surface skin.  The difficulty with these models is
the much more massive disk required. The gross model geometry of
the disk would not change, still crossing the line of sight to
the center at the same radius.  However, the column density would be much greater at the crossing
radius than determined by the X-ray observations. 

Three other reservoirs for absorbing gas are also conceivable: an edge-on
molecular torus at much larger radii than the ``maser disk'' (as is often
assumed in the AGN unification paradigm), an outflow \citep[e.g.,][]{elvis00}, and an  
extension of the disk to much smaller radii.   We argue
against the presence of a massive torus because the CO
distribution in \4258\ is not centrally peaked \citep[][and references
therein]{plante91}. We argue against outflow because \4258\
is a low-luminosity AGN and the possibility of it driving a massive
outflow seems remote. In addition, radio and optical imaging show a 
well-collimated flow \citep{herrnstein97, cecil00} that lies well away from the
line of sight to the central engine.  The possibility of absorption arising
in a disk-like structure at small radii ($\ll 0.14$~pc) exists, but is
unlikely given the three-dimensional geometry of the maser disk.  We note
that the disk inclination is $\sim 82^\circ$ at a radius of 0.14~pc.  For an
$m=1$ warping mode, the inclination is more face-on at smaller radii (down to $\sim 74^\circ$). 
Hence, disk material at small radii cannot lie along the line of sight to the central
engine unless the disk is significantly thickened ($h/r>0.3$). If the
absorbing material is neutral and cold, this raises the question of what
process provides vertical support. The problem of support is mitigated if
the disk is very hot, but the X-ray spectrum of \4258\ does not show signs
of very hot gas.

\subsection {Fe K$\alpha$ Emission Line}

We do not detect Fe K$\alpha$ line emission in any of the \XMM\ epochs
or in any of the \Chandra\ epochs. The \XMM\
observation on 2001 May 6 puts a stringent 90\% upper limit to the
equivalent width of $<49$ eV. In light of the last detection of the
line with \ASCA\ at epoch 1999 May 15 (Table~2), \citet{pietsch02} and
\citet{rn03} concluded the line is variable, fading on a time scale of
$<10$ months.  The apparent disappearance, which now extends through
May 2002, is difficult to explain, assuming the earlier detections with
\ASCA\ (at multiple epochs) were robust.

The accretion disk is the most prominent reservoir of cold material
known in close proximity to the central engine.   Assuming that
the line originates in the disk, its weakness is consistent with the
inferred  relatively low column density of the disk ($\sim
10^{23}$~cm$^{-2}$), which enables transmission of incident hard X-rays
(i.e., 6.4~keV) virtually unattenuated.  As a result, reflection
should be negligible (see also Fiore et al. 2001), and
\citet{ghisellini94} have argued that at such low column densities, 
equivalent widths for line fluorescence should be $\ll 100$~eV.

The weakness of the line is also consistent with the disk subtending
a small solid angle ($\Omega/2\pi\ll1$) as seen from the source of hard
X-ray continuum near the disk center.  This would be the case if 
\4258\ were to host an advection-dominated accretion flow (ADAF). 
\citet{herrnstein97} measured the angular structure of radio continuum
emission in the vicinity of the disk center, and showed that any ADAF
must be compact, $< 100$ Schwarzschild radii.  No cold
material would extend toward the center, and as a result the disk solid
angle would be small.   Although the presence of an ADAF could explain
the weakness of Fe line emission, our earlier argument that a large mass
accretion rate and thick disk are inconsistent with the observed X-ray
absorbing column remains valid.  In addition, we note that \citet{blandford99}
suggest that ADAFs should be accompanied by massive winds (thus
depleting the accretion flow), and \citet{fiore01} argue that such
winds are inconsistent with the broadband spectral energy distribution
of \4258. In place of an ADAF as the source of hard X-ray continuum
emission, a disk corona remains a possibility.  Although in this case the
disk could extend inward far enough that it would subtend a large solid
angle ($\Omega/2\pi \approx 1$), the relative (optical) thinness of the
disk alone could explain the weakness of any Fe K$\alpha$ emission.  It
also remains possible that the line emission arises from scattering,
fluorescence, or reflection by material that lies away from the
disk, but apart from the well-collimated radio synchrotron jet
\citep{herrnstein97}, none is known.

Past time variability of the line is more difficult to explain: light
travel time arguments applied to the fading suggest that the 
line-emitting region is $< 0.5$~pc in size, corresponding to the
approximate outer diameter of the maser disk.  The changes in the
line may be the result of changes in the disk across its surface
area or changes in the central illumination.  However, if both X-ray and radio
(maser) line emission are the result of irradiation of the disk
surface, then it is notable that the brightness and velocity extent of
H$_2$O maser emission have not changed substantially since monitoring
began in 1992
\citep[e.g.,][]{nakai93, bragg00, argon05}. This would suggest that the disk 
(at 0.14--0.28~pc radii) and its source of illumination have remained relatively unchanged
over the period that the Fe line has faded.  However, considering the
fact that both maser emission and Fe line emission may be influenced by
other effects (e.g., propagation along the disk, gas heating and
cooling rates), coordinated~ X-ray and radio monitoring over time scales of
$\la 1$~day to $\sim 6$~months is needed to more firmly detect the line
and to test proposed emission and variability mechanisms.

\subsection {Iron Absorption Lines}
Time-variable Fe line absorption (at 6.4 and 6.9 keV) has been
reported from a study of one (2000 April 28) of the \Chandra\
observations by Young \& Wilson (2004). In our analysis 
we find no evidence of absorption lines, neither in the \XMM\
nor in the \Chandra\ spectra.  Because our findings differ from what was
previously reported, we tested our results by retracing, to the best
of our knowledge, these authors' data-analysis steps.  We repeated the
analysis using several binning schemes for the extracted spectra
(including a minimum of 15 counts per bin as used by the previous
authors), we inspected the data split in two equal time intervals, and we
applied pile-up models with various choices of parameters.
Figures 8 to 10 illustrate our results: no significant absorption
feature can be inferred from our data and best-fit models.  We can
only speculate about possible origins for the discrepancy: differences
in calibration, or slight differences in the data preparation, 
treatment of the background, or the modeling of pile-up. We note
that the 2000 April 28 epoch is the one most affected by
photon pile-up at a level of $\sim30\%$ (see Table 3 and \S 2.3.1).

\section{Summary and Conclusions}

In this paper we have analyzed X-ray (0.3--10~keV) observations of
\4258~ obtained with the \XMM\ and \Chandra\ observatories over $\sim 2$ years. 
Including earlier observations by \ASCA\ and \SAX, 
we present for the first time a new nine-year time-series of models fit to the
X-ray data from \4258.

Spectral and time variability analyses lead to the following main results:

(1) The \XMM\ and \Chandra\ spectra of \4258\ data are well fit by a model
consisting of several components: a partially absorbed, hard ($>2$ keV) power law, 
a soft thermal plasma, and a soft power law.  
The soft emission, some of which arises $\la 70$~pc from the
central engine, does not vary appreciably from observation to observation.

(2) From \XMM\ data, which are free of systematics such as pile-up, we observe 
long-term time variability in the source count rate 
and absorbed flux
over time scales of $\sim 6$ months. No evidence for 
variability, however, is detected  within our individual $\sim 3$~hour integrations.

(3) From \XMM\ data we detect a $\sim 60\%$ increase in $N_H$~ over 
$\sim 5$~months, returning to a high level not
reported since \ASCA\ observations in 1993, $N_H \approx
1.3 \times 10^{23}$~cm$^{-2}$.

(4) Changes in $N_H$ and $f_X$ are not correlated, which
indicates intrinsic variability of the central engine that is in one
case 30\% over 19~days (5--10~keV). We note that two of the largest
estimates of unabsorbed luminosity are associated with the lowest
estimates of $N_H$, and we speculate that reductions in $L_X$ might
affect the ionization state of the absorber.

(5) The geometry and orientation of the
accretion disk in \4258\ is well known from interferometric mapping of
maser emission that arises in the accretion disk. We infer that the
disk is somewhat inhomogeneous, based on observed time variability in
$N_H$ and an assumption that the absorbing gas lies in the disk.
Using the rotation curve of the disk, we estimate that the absorbing gas
may lie at the same radius as where the disk is (independently)
believed to cross our line of sight to the central engine (because the
disk is warped) and that the inhomogeneities are 
$\sim 10^{15}$~cm in size.

(6) We do not detect Fe K$\alpha$ line emission in any of the \XMM\  or
\Chandra\ epochs, thus extending the ``disappearance'' of the line
from the last~ \ASCA\ detection in 1999 May to 2002 May.   The
inferred line emission region is comparable in size to the maser disk. 
If the line arises from the disk (e.g., by fluorescence), then it
is difficult to understand the variability  because the maser emission
has not changed substantially.

(7) We do not observe evidence for absorption lines in any \XMM\
or \Chandra\ spectra.

Our inference that the accretion disk is inhomogeneous at radii
of $\sim 0.3$~pc raises the question of what mechanism is responsible.
Assessment of the Toomre Q-parameter at this radius suggests that
self-gravity is unimportant except at the highest possible densities
supportive of maser action
\citep[cf.][]{maoz95}.  
Magnetic fields may play a role in confining inhomogeneities, but the
field strength and configuration are not known
\citep{herrnstein98,modjaz05}.  
While the confinement mechanism is not determined, we note that
detection of inhomogeneities in the disk empirically supports
techniques to estimate a distance to \4258\ from maser proper motions
and centripetal accelerations, because these rely on the apparent
maser dynamics tracking the rotation of disk material \citep[][and
references therein]{herrnstein99}.

\acknowledgments
We thank Tom Aldcroft, Jeremy Drake, Martin Elvis, Greg Madejski,
Guido Risaliti, and Aneta Siemiginowska for helpful discussions and
insightful comments. AF and LJG are grateful to Martin Elvis for
fostering their collaboration.  This work has been supported by NASA
GO grants NAG5-10194 and NAG5-11689, and by NASA science grant
NAS8-39073 to the Chandra X-Ray Center.

%\clearpage

%%%%%%%%%%%%%%%%%%%%%%%%%%%%%%%%%%%%%%%%%%%%%%%%%%%%%%

%numbers come from pn{/mos1/mos2}_{evt2/evt2_filt}

\clearpage

\begin{deluxetable}{llccc}
\tablecaption{\XMM\ EPIC OBSERVATIONS}
\tablehead{
\colhead{Observation} &
\colhead{Instrument}  &
\colhead{Exposure} &
\colhead{Filtered} &
\colhead{Average}
\\
\colhead{Date} &
\colhead{}  &
\colhead{Time} &
\colhead{Exp. Time\tablenotemark{a}} &
\colhead{Count rate}
\\
\colhead{} & 
\colhead{} & 
\colhead{[s]} & 
\colhead{[s]} & 
\colhead{[c s$^{-1}$]} 
}
\startdata

2000-12-08\tablenotemark{b}  &PN   &14442 &14442 &0.734$\pm$0.007\\
            & MOS1 &20250 &20250 &0.224$\pm$0.004\\
            & MOS2 &20261 &20261 &0.228$\pm$0.003\\
\\
2001-05-06 &PN   & 8597 & 8597 &0.623$\pm$0.009\\
                  & MOS1 &11998 &11998  &0.180$\pm$0.004\\
                  & MOS2 &12001 &11985 &0.182$\pm$0.004\\
\\
2001-06-17 &PN    &9055 & 9055 &0.568$\pm$0.008\\
                    &MOS1 &12438 &12438 &0.166$\pm$0.004\\
                    &MOS2 &12441 &12441 &0.166$\pm$0.004\\
\\
2001-12-17 & PN    &7917 &6850  &0.523$\pm$0.009\\
                    &MOS1 &12459 &10783 &0.147$\pm$0.004\\ 
                   &MOS2 &12498 &10584 &0.147$\pm$0.004\\ 
\\
2002-05-22 &PN   &12089 &11755 &0.420$\pm$0.007\\
                    &MOS1 &15664 &15461 &0.113$\pm$0.003\\
                   &MOS2 &15664 &15664 &0.122$\pm$0.003\\

\enddata
\tablenotetext{a}{Total ``good'' exposure time after filtering out times when the ratio of source counts to background count was less than 5.}
\tablenotetext{b}{This observation was also analyzed and published by Pietsch \& Read (2002).}
\end{deluxetable}

\clearpage
\thispagestyle{empty}

\begin{deluxetable}{lllllllllllllll}
\rotate
\tabletypesize{\scriptsize}
\tablenum{2}
\tablecolumns{15}
\tablewidth{0pt}
\tablecaption{SPECTRAL DATA FOR \4258\ \tablenotemark{(a)}}
\tablehead{
\colhead{Observatory} & 
\colhead{Date} & 
\colhead{$kT$} & 
\colhead{$f_{0.5-2}$\tablenotemark{(b)}} & 
\colhead{$L_{0.5-2}$\tablenotemark{(c)}} & 
\colhead{$N_{H}$\tablenotemark{(d)}} &
\colhead{$\Gamma_{pl}$} &
\colhead{$f_{5-10}$\tablenotemark{(b)}} &
\colhead{$L_{5-10}$\tablenotemark{(c)}} &
\colhead{$f_{2-10}$\tablenotemark{(b)}} & 
\colhead{$L_{2-10}$\tablenotemark{(c)}} & 
\colhead{$E_{line}$} & 
\colhead{$EW_{line}$} &
\colhead{$f_{line}$\tablenotemark{(b)}} &
\colhead{Ref.}         
\\
\colhead{} & 
\colhead{} & 
\colhead{[keV]} &
\colhead{[$10^{-12}$]}&
\colhead{[$10^{38}$]}& 
\colhead{[$10^{22}$ cm$^{-2}$]} & 
\colhead{} & 
\colhead{[$10^{-12}]$} &
\colhead{[$10^{40}$]}&
\colhead{[$10^{-12}$]}&
\colhead{[$10^{40}$]}&
\colhead{[keV]} & 
\colhead{[eV]} & 
\colhead{[$10^{-13}$]}&
\colhead{}              
\\
}
\startdata
\it ASCA &1993 May 05 &$0.5\pm0.2$\tablenotemark{(e,f)} & & &$15\pm2$ &$1.78\pm0.29$ & & & &4.2 &$6.5\pm0.2$ &$250\pm100$ & &1\\
&1993 May 05 & & & &$13.6^{+2.1}_{-2.2}$ &$1.78^{+0.22}_{-0.26}$ &5.1 &4 & & & & & &2
\vspace{0.05in}\\

\it ASCA &1996 May 23 & & & &$9.2\pm0.9$ &$1.71^{+0.18}_{-0.17}$ &8.3 &6.1& & & & & &2\\
\it ASCA &1996 Jun 5  & & & &$8.8\pm^{+0.7}_{-0.6}$ &$1.83\pm0.13$ &8.8 &6.4 & & & & & &2\\
\it ASCA &1996 Dec 18 & & & &$9.7\pm0.8$ &$1.87\pm0.15$ &9.5 &6.9 & & & & & &2\\
{\it Beppo}-SAX &1998 Dec 19 &$0.6\pm0.1$ &1.6\tablenotemark{(e)} & &$9.4\pm1.2$ &$2.11\pm0.14$ &5.2 &3.8 &8 &10 &$6.57\pm0.20$ &$85\pm65$ & &3
\vspace{0.05in}\\
% data from Reynolds et al. 2000
\it ASCA &1999 May 15 &$0.47^{+0.03}_{-0.09}$ & & &$8.2\pm0.9$ &$1.79^{+0.31}_{-0.11}$ & & & &5.8 & & & &2\tablenotemark{(g)}\\ %model F (best fit)
&1999 May 15 &$0.36^{+0.03}_{-0.02}$  & & &$9.5^{+2.1}_{-0.9}$ &$1.86^{+0.40}_{-0.13}$ &4.0 &2.9 & & &$6.45^{+0.10}_{-0.07}$ &$107^{+42}_{-37}$ & &2\tablenotemark{(h)} %model G (=model E +Gaussian)
%%%%%%%%%%%%%%%%%%%%%%%%%%%%%%%CHANDRA AND XMM DATA BELOW%%%%%%%%%%%%%%%%%%%%%%%%%%%%%%%
\vspace{0.05in}\\
% OBSID 349
\it Chandra &2000 Mar 08 & & & &$7.2\pm1.8$ &$1.4\pm0.5$ & & &14.4 &13.2 & & & &4\\
&2000 Mar 08 &N/A &N/A &N/A &$6.9^{+2.2}_{-1.8}$ &$1.3\pm{0.6}$ & & & &12 &6.4&$<887$ & &5\\
%with bin=10
&2000 Mar 08 &N/A &N/A &N/A &$6.0^{+1.8}_{-1.4}$ &$1.0^{+0.5}_{-0.4}$ &$15.9$ &11.0 &$21.2$ &$17.4$ &6.4 &$<369$ & $<12$ &6 %final
\vspace{0.05in}\\
%%%%%%%%%%%%%%%%%%%%%%%%%%%%%%%
%OBSID 350 (my values are for bin10 and chi dvar)
\it Chandra &2000 Apr 17 &$1.3^{+\infty}_{-0.9}$\tablenotemark{(f)} & &1.6 &$7.2^{+0.7}_{-0.4}$ &$1.5^{+0.1}_{-0.0}$ & & & &13 & & & &5\\
% my runs 
&2000 Apr 17\tablenotemark{(i)} &$0.5\pm0.1$ &0.07 &1.7 &$6.7^{+1.0}_{-0.9}$ &$1.4\pm0.3$ &8.3 &5.8 &11.8 &10.6 &6.4 &$<61$ &$<1.2$ &6 \\ %alpha free, f free     => redchi=1.18398, pileup fraction:0.277811
\vspace{0.05in}\\
%%%%%%%%%%%%%%%%%%%%%%%%%%%%%%%
XMM-{\it Newton} &2000 Dec 08 & & & &$8.0\pm0.4$ &$1.64\pm0.08$ & & &7.6 &7.5 &6.45 &$<40$ &  & 4\\
&2000 Dec 08 &$0.6\pm0.03$ &$0.2$ & $14.6$ &$8.6\pm0.4$ &$1.7\pm0.1$ &$6.1$ &4.4 &$9.0$ &$9.1$ &6.4 &$<69$ &$<1.0$ &6 %newdatanew
%%%%%%%%%%%%%%%%%%%%%%%%%%%%%%%
\vspace{0.05in}\\
XMM-{\it Newton} &2001 May 06 &$0.5^{+0.06}_{-0.05}$ &$0.2$ &$15.3$ &$9.8\pm0.7$ &$1.9\pm0.1$ &$4.8$ &3.5 &$7.1$ &$7.8$ &6.4 &$<49$ &$<0.6$ &6 %20010722 newdatanew
%%%%%%%%%%%%%%%%%%%%%%%%%%%%%%%
\vspace{0.05in}\\
%OBSID 1618 (my values are for bin10 and chidvar)
\it Chandra &2001 May 28 &$1.3^{+\infty}_{-0.8}$\tablenotemark{(f)} & & 1.4 &$6.5^{+0.6}_{-0.3}$ &$1.5\pm0.1$ & & & &7.9 &6.4  &$<132$ & &5\\
% my runs 
&2001 May 28\tablenotemark{(i)} &$0.6^{+\infty}_{-0.18}$ &0.06 &1.3 &$6.0\pm0.7$ &$1.4^{+0.3}_{-0.2}$ &6.6 &4.5 &9.4 &8.1 &6.4 &$<191$ &$<2.8$ &6 \\ %alpha free, f free, redchisq=1.28603, pileup=0.161134
%%%%%%%%%%%%%%%%%%%%%%%%%%%%%%%
\vspace{0.05in}\\
%OBSID 2340
\it Chandra &2001 May 29 &$1.0^{+0.6}_{-0.7}$\tablenotemark{(f)} & &1.5 &$6.8^{+1.2}_{-1.0}$ &$1.4\pm{0.3}$ & & & &7.4 &6.4 &$<94$ & &5\\
% with bin=10
&2001 May 29 &$0.8\pm0.2$ &0.06 &1.5 &$5.9^{+0.8}_{-0.7}$ &$1.3\pm0.2$ &6.2 &4.3 &8.8 &7.6 &6.4 &$<100$ &$<1.4$ &6 %final
%%%%%%%%%%%%%%%%%%%%%%%%%%%%%%%
\vspace{0.05 in}\\
XMM-{\it Newton} &2001 Jun 17 &$0.5^\pm0.05$ &$0.2$ &$14.1$ &$8.5^{+0.7}_{-0.6}$ &$1.8\pm0.1$ &$4.3$ &3.1 &$6.4$ &$6.5$ &$6.3^{+0.07}_{-0.08}$\tablenotemark{(j)} &$66\pm43$\tablenotemark{(j)} &0.7 &6 \\ %20010726 newdatanew
%%%%%%%%%%%%%%%%%%%%%%%%%%%%%%%
XMM-{\it Newton} &2001 Dec 17 &$0.5^{+0.06}_{-0.12}$ &$0.3$ &$16.4$ &$8.4^{+1.0}_{-0.9}$ &$1.5^{+0.2}_{-0.1}$ &$4.1$ &2.9 &$5.8$ &$5.5$ &6.4 &$<67$ &$<0.7$ &6 \\% newdatanew
%%%%%%%%%%%%%%%%%%%%%%%%%%%%%%%
XMM-{\it Newton} &2002 May 22 &$0.5\pm0.04$ &$0.2$ &$14.9$ &$13.2^{+1.4}_{-1.3}$ &$1.5\pm0.2$ &$3.7$ &2.8 &$4.9$ &$5.4$ &6.4 &$<100$ &$<0.9$ &6 \\% newdatanew
%%%%%%%%%%%%%%%%%%%%%%%%%%%%%%%
\enddata
\hspace{-0.6in} References-- 1: Makishima et al. (1994), 2: Reynolds et al. (2000), 3: Fiore et al. (2001), 
4: Pietsch and Read (2002), 5: Young and Wilson (2004), 6: this work\\ 
\vspace{-0.15in}
\tablenotetext{(a)} {Results reported here are from model fits performed on the 0.3--10 keV (\XMM\ ) and 0.5--10 keV(\Chandra\ ) energy ranges. 
The models comprise several components.  For {\it Chandra}: Galactic absorption ($N_H(gal)=1.19\times10^{20}$ cm$^{-2}$, frozen in the fits), intrinsic
absorption ($N_H$), low energy thermal plasma component ($kT$), power law ($\Gamma_{pl}$), and Gaussian Fe K-line ($E_{line}$, $W_{line}$). 
For \XMM\ an additional soft power law absorbed by $N_H(gal)$ has also been added. The slope of this component remains almost constant in all observation with an average $<\Gamma> = 1.6\pm0.2$. The energy of the  Fe K-line was frozen at 6.4 keV for all observations except for the \XMM\ 2001-06-17. All errors are quoted at 90\% confidence.}
\tablenotetext{(b)} {Absorbed fluxes (ergs cm$^2$ s$^{-1}$) in the 0.5--2~keV, 5--10~keV, and 2--10~keV bands and Fe K-line flux or upper limit. For \SAX\ data, $f_{5-10}$ is derived via PIMMS assuming the published values of $f_{2-10}$, $N_H$, and $\Gamma_pl$}.
\tablenotetext{(c)} {Absorption-corrected luminosities (ergs s$^{-1}$) in the 0.5--2~keV, 5--10~keV and 2--10~keV bands, 
assuming a distance of 7.2 Mpc. Both Galactic and intrinsic absorption have been removed. For \ASCA\ and \SAX\ data $L_{5-10}$ is derived via PIMMS assuming the values of $f_{5-10}$, $N_H$, and $\Gamma_pl$.}
\tablenotetext{(d)} {Intrinsic equivalent hydrogen column density.}
\tablenotetext{(e)} {0.1--2.4~keV energy band.}
\tablenotetext{(f)} {Bremsstrahlung temperature.}
\tablenotetext{(g)} {Model F (best fit) in Table 1 of Reynolds et al. (2000).}
\tablenotetext{(h)} {Model G (with added Gaussian) in table 1 of Reynolds et al. (2000). 
Note that there are some typos in paragraph 3.2 of Reynolds et al. (2000): model D, model E, and model F should read 
model E, model F, and model G, respectively.}
\tablenotetext{(i)} {A pile-up model with $\alpha$ and $f$ free to vary was added to the model components. See text and Table 3 for details.}
\tablenotetext{(j)} {Values obtained by leaving $E_{line}$ free to vary. The fit is not significantly improved by the addition of a Gaussian line
(reduced $\chi^2$=1.05 for 159 degrees of freedom without a line, reduced $\chi^2$=1.02 for 156 degrees of freedom with a line).}
\end{deluxetable} 

\clearpage

\begin{deluxetable}{llllllllllcl}
%\rotate
\setlength{\tabcolsep}{0.02in}
\tabletypesize{\tiny}
\tablenum{3}
\tablecolumns{12}
\tablewidth{0pt}
\tablecaption{COMPARISON OF PILE-UP MODELING FOR CHANDRA OBSERVATIONS\tablenotemark{(a)}}
\tablehead{
\colhead{Date} & 
\colhead{$N_{H}$} &
\colhead{$\Gamma_{pl}$} &
\colhead{$f_{5-10}$} &
\colhead{$L_{5-10}$} &
\colhead{$f_{2-10}$} & 
\colhead{$L_{2-10}$} & 
\colhead{$W_{line}$} &
\multicolumn{3}{c}{Pile-up Model}   &
\colhead{$\chi^2$/dof}  
\\
\colhead{} & 
\colhead{[$10^{22}$ cm$^{-2}$]} & 
\colhead{} & 
\colhead{[$10^{-12}]$} &
\colhead{[$10^{40}$]}&
\colhead{[$10^{-12}$]}&
\colhead{[$10^{40}$]}&
\colhead{eV} & 
\colhead{$\alpha$\tablenotemark{(b)}} &  
\colhead{$f$\tablenotemark{(c)}} &  
\colhead{\%\tablenotemark{(d)}} & 
\colhead{} 
\\
}
\startdata
2000 Apr 17 &$6.7^{+1.0}_{-0.9}$ &$1.4\pm0.3$ &8.3 &5.8 &11.8 &10.6 &$<61$ &0.65 &0.90 &28 &1.2=66.8/59 \\ %alpha free, f free     => redchi=1.18398, pileup fraction:0.277811
            &$7.3^{+1.2}_{-1.0}$ &$1.8^{+0.4}_{-0.3}$ &8.0 &5.7 &12.2 &12.1 &$<76$  &0.60 &0.95\tablenotemark{(e)} &30 &1.4=85.8/60  \\ %alpha free, f frozen => redchi=1.43023, pileup fraction: 0.308118
            &$7.3^{+1.2}_{-1.0}$ &$1.7\pm0.4$ &9.4 &6.6 &14.1 &13.7 &$<73$  &0.5\tablenotemark{(e)} &0.95\tablenotemark{(e)} &29 &1.5=89.7/61 \\ %alpha frozen, f frozen   => redchi=1.47098 pileup fraction: 0.294481
2000 Apr 17\tablenotemark{(f)} &$7.2^{+0.7}_{-0.4}$ &$1.5^{+0.1}_{-0.0}$ & & & &13 &  &0.41 &0.95\tablenotemark{(e)} & &1.1=113/103 \\ %alpha frozen, f frozen   => redchi=1.47098 pileup fraction: 0.294481
\vspace{0.05in}\\
2001 May 28 &$6.0\pm0.7$ &$1.4^{+0.3}_{-0.2}$ &6.6 &4.5 &9.4 &8.1 &$<191$   &0.43 &0.91 &16 &1.3=75.9/59 \\ %alpha free, f free, redchisq=1.28603, pileup=0.161134
            &$6.2^{+0.8}_{-0.7}$ &$1.5\pm0.3$ &8.9 &6.2 &13.0 &11.5 &$<203$ &0.31 &0.95\tablenotemark{(e)} &17 &1.3=78.1/60   \\ %alpha free, f frozen  redchisq=1.30203, pileup=0.16781
            &$6.6\pm0.8$ &$1.7\pm0.3$ &5.8 &4.0 &8.8 &8.3 &$<234$  &0.5\tablenotemark{(e)} &0.95\tablenotemark{(e)} &20 &1.5=88.8/61 \\ %alpha frozen, f frozen  redchisq=1.45632, pileup=0.19696
2001 May 28\tablenotemark{(f)} &$6.5^{+0.6}_{-0.3}$ &$1.5^{+0.1}_{-0.0}$ & & & &7.9 &$<132$  &0.43 &0.95\tablenotemark{(e)} & &1.1=158/143 \\ %alpha frozen, f frozen  redchisq=1.45632, pileup=0.196961
\enddata
\\
\vspace{-0.15in}
\tablenotetext{(a)} {Model components as in Table 2.} 
\tablenotetext{(b)}{Probability of obtaining a good event grade when two photons pile together.}
\tablenotetext{(c)}{Fraction of flux falling into the pile-up region.}
\tablenotetext{(d)}{Percentage of observed counts containing two or more photons piled together.}
\tablenotetext{(e)}{Frozen parameter.}
\tablenotetext{(f)}{Best-fit parameters from Young \& Wilson (2004).}

\end{deluxetable} 

\clearpage

\newpage
\begin{figure}
\epsscale{1.0}
\plotone{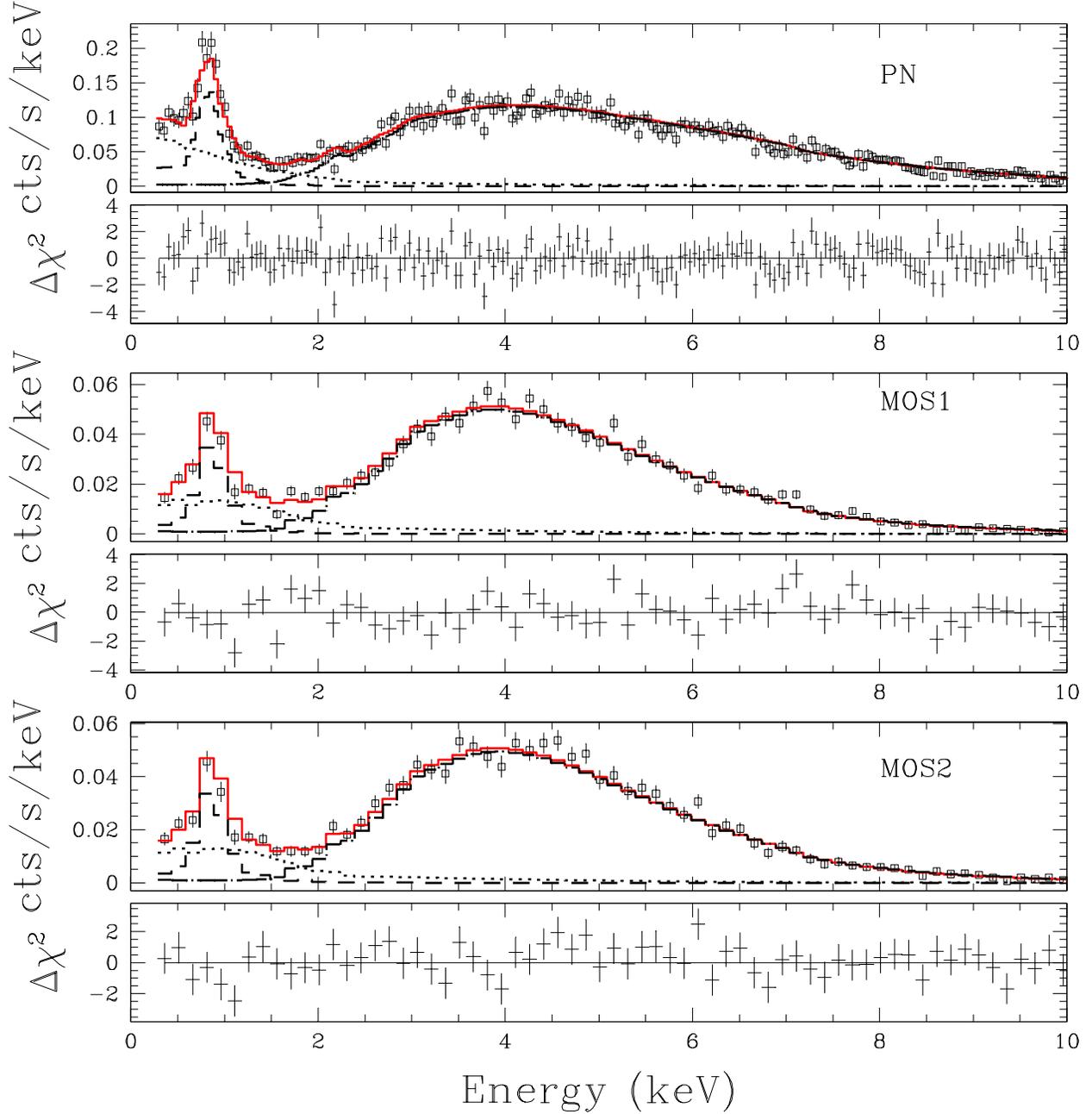}
\caption{\XMM\ EPIC spectra of the nucleus (15\arcsec\ extraction radius) of \4258\ for the observation of 2000-12-08. Data were ``grouped'' by 10 pulse-invariant (PI) channels.
PN, MOS1, and MOS2 data were fitted simultaneously with a model consisting of 2 power laws (dotted and dot-dashed lines) plus a thermal spectrum (dashed line) absorbed by intrinsic and Galactic gas (N$_H$) (see text). The best fit (solid line) and residuals (in units of $\sigma$) are shown for all instruments. 
These are the same data presented by Pietsch \& Read (2002) (note that their Figures 2 and 5 report units of counts~cm$^{-2}$~s$^{-1}$~keV$^{-1}$ while they should read counts~s$^{-1}$~keV$^{-1}$, as in our plots, given that the spectra are convolved with the instrumental ancillary response function).}
\label{f:spectrum0_xmm}
\end{figure}

\newpage
\begin{figure}
\epsscale{1.0}
\plotone{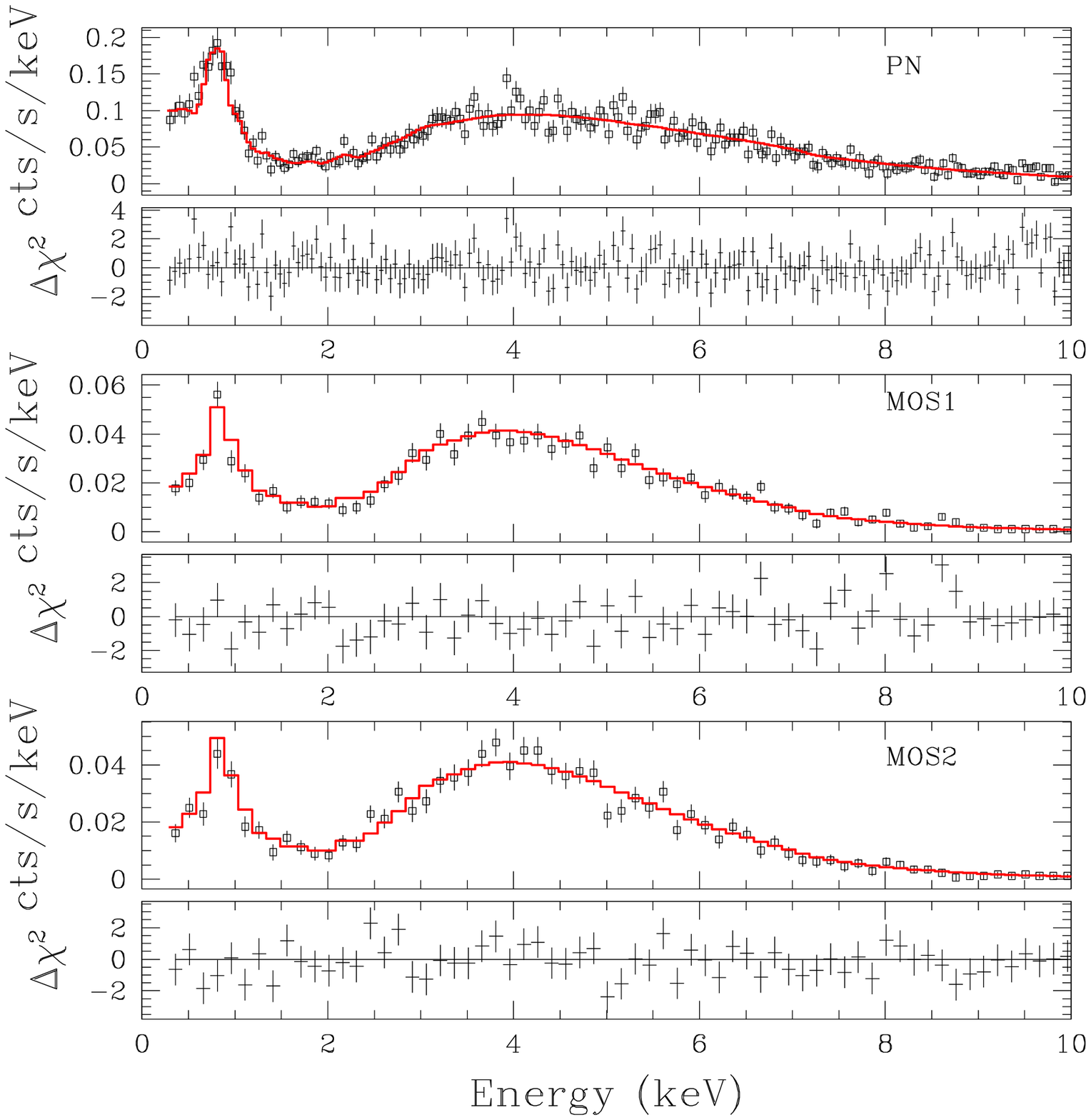}
\caption{Same as Fig.~\ref{f:spectrum0_xmm} for the \XMM\ observation of 2001-05-06.}
\label{f:spectrum1_xmm}
\end{figure}

\newpage
\begin{figure}
\epsscale{1.0}
\plotone{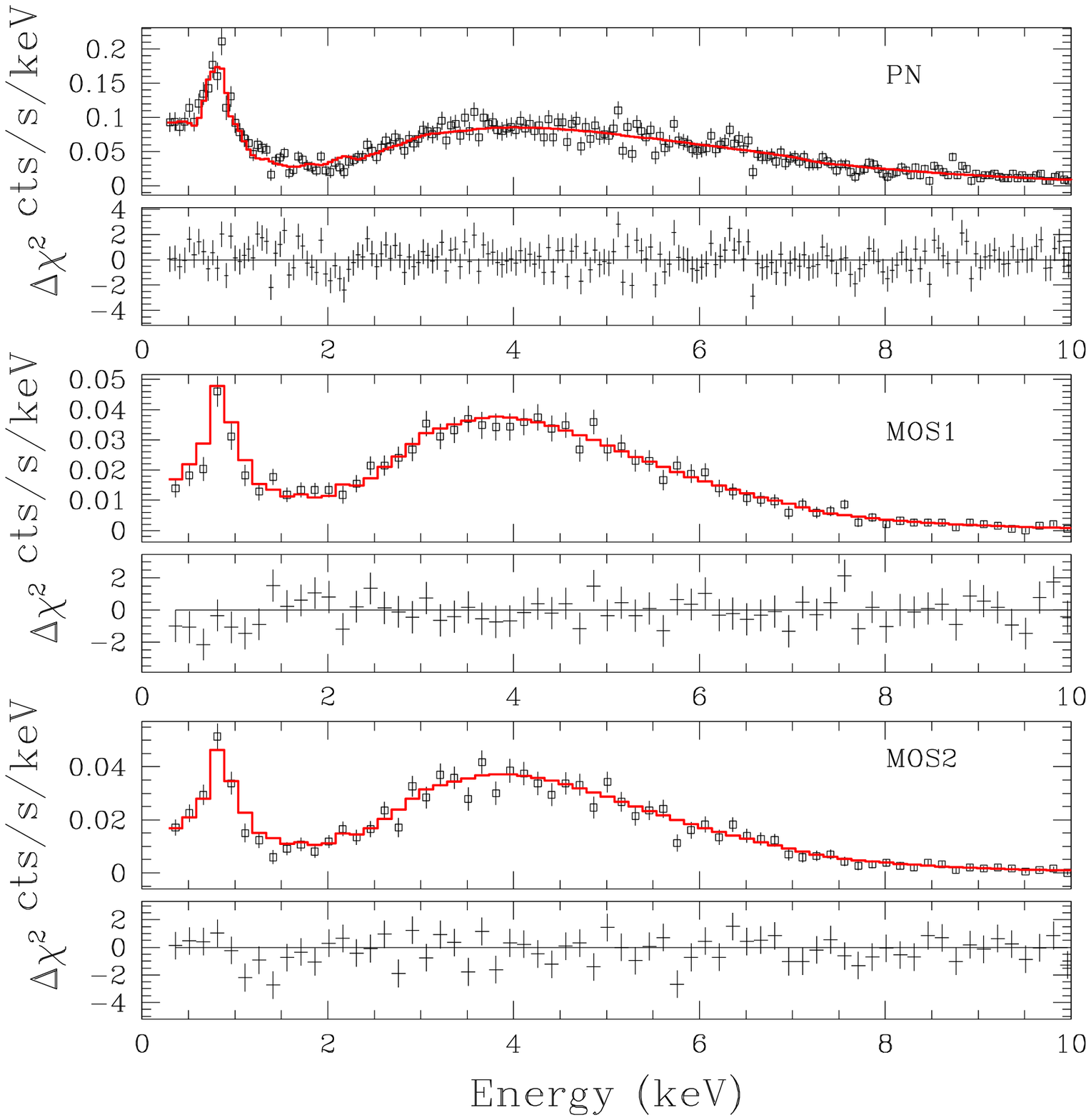}
\caption{Same as Fig.~\ref{f:spectrum0_xmm}  for the \XMM\ observation of 2001-06-17.}
\label{f:spectrum2_xmm}
\end{figure}

\newpage
\begin{figure}
\epsscale{1.0}
\plotone{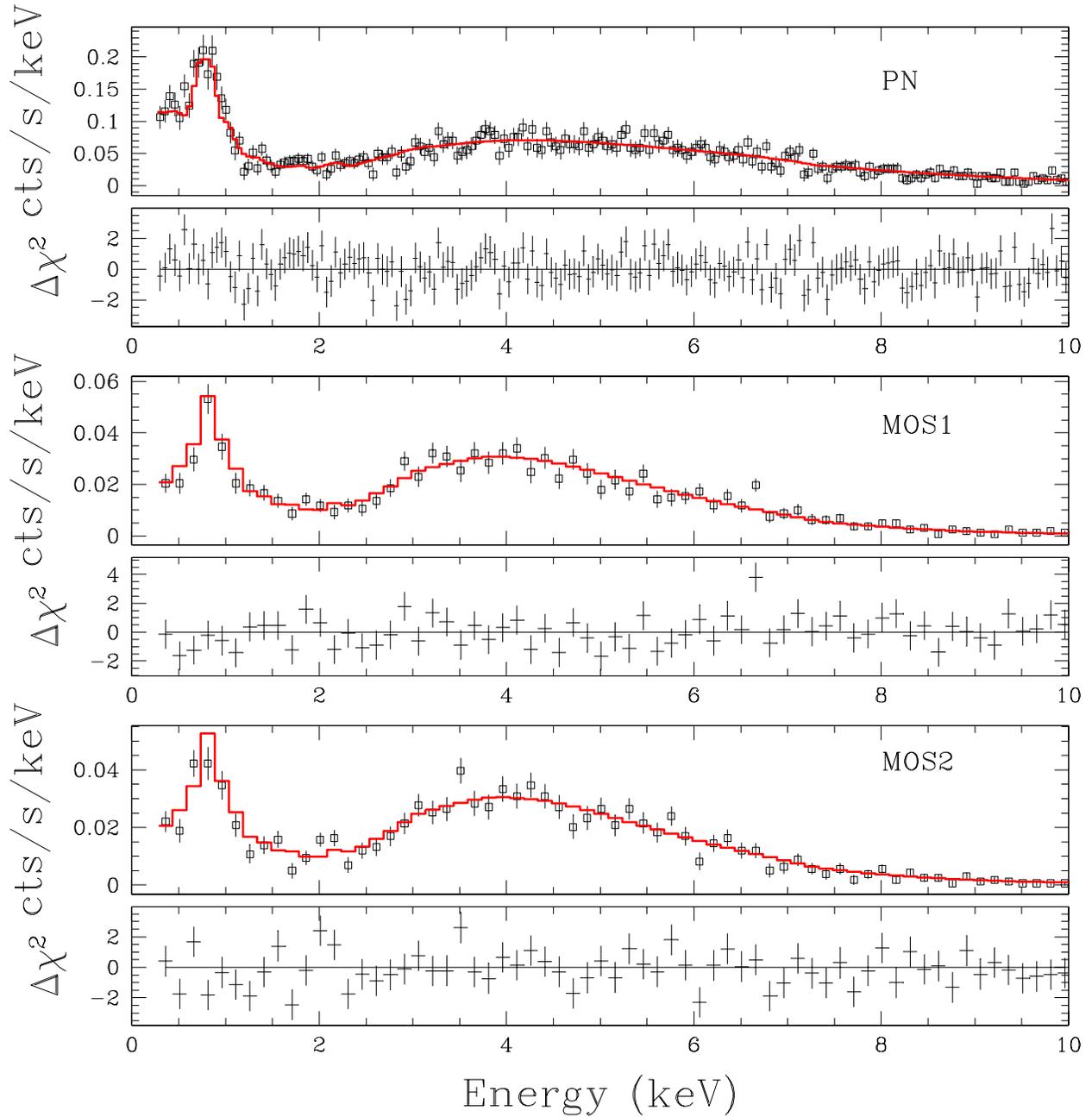}
\caption{Same as Fig.~\ref{f:spectrum0_xmm} for the \XMM\ observation of 2001-12-17.}
\label{f:spectrum3_xmm}
\end{figure}

\newpage
\begin{figure}
\epsscale{1.0}
\plotone{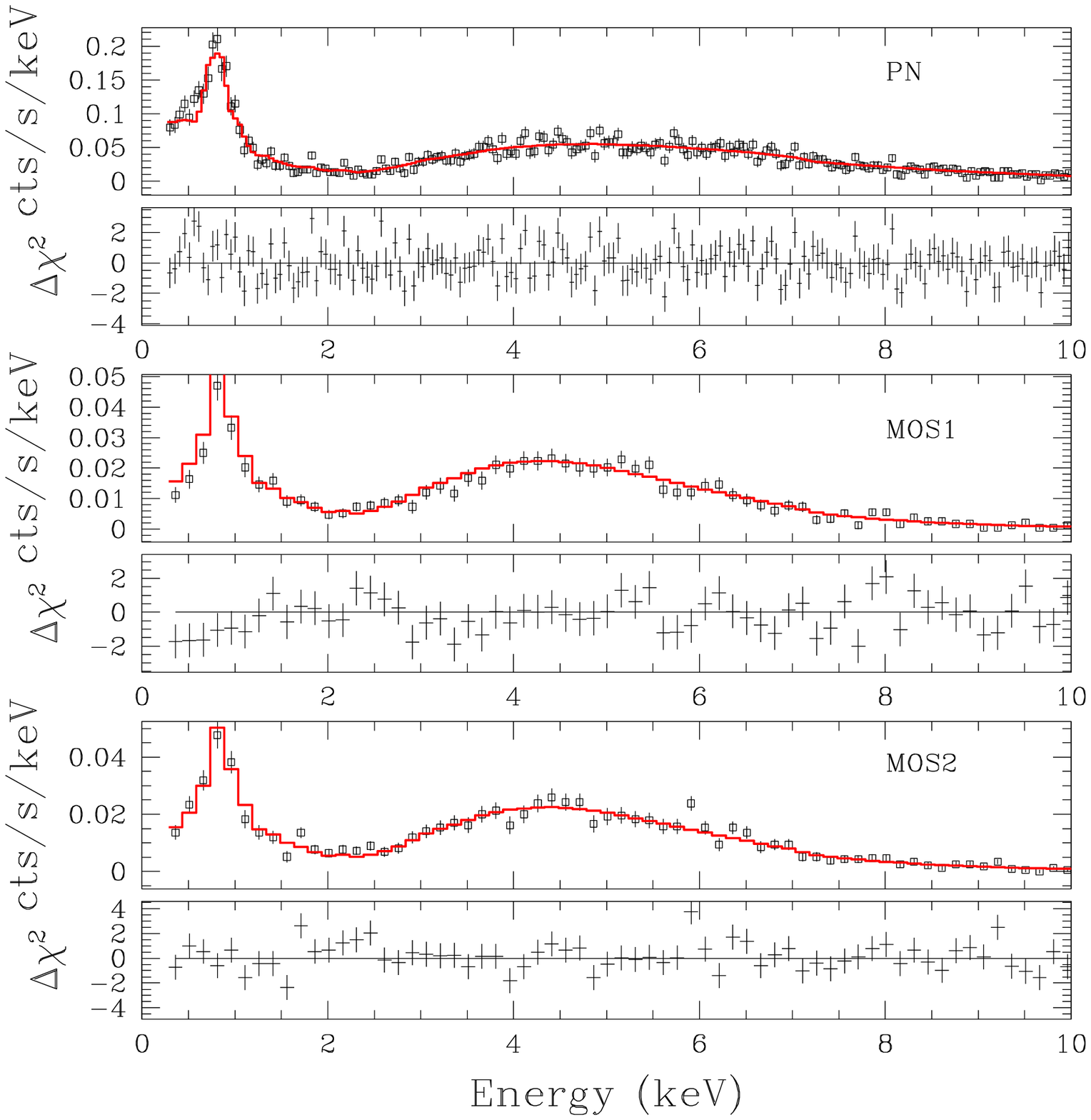}
\caption{Same as Fig.~\ref{f:spectrum0_xmm} for the \XMM\ observation of 2002-05-22.}
\label{f:spectrum4_xmm}
\end{figure}

\newpage
\begin{figure}
\epsscale{1.0}
\plotone{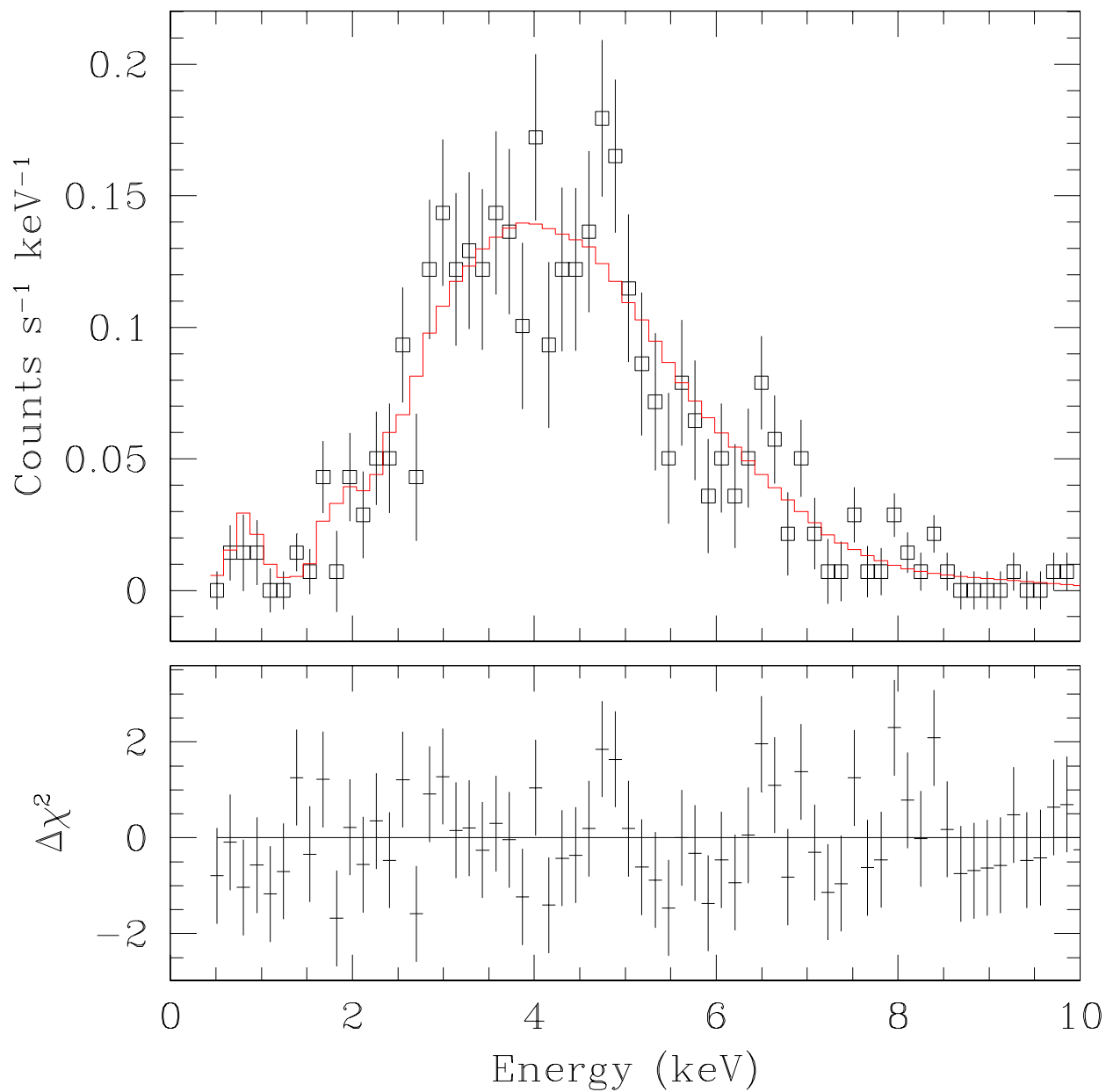}
\caption{\Chandra\ spectrum of the nucleus (2\arcsec\ extraction radius) of \4258\ for the observation of 2000-03-08. Data were ``grouped'' by 10 PI channels.
The data were fitted with a model consisting of a power law plus a thermal component absorbed by intrinsic and Galactic gas (N$_H$) (see text). These are the same data presented by Pietsch \& Read (2002) (the final note of the caption in Fig.~\ref{f:spectrum0_xmm} applies).}
\label{f:spectrum1_chandra}
\end{figure}

\newpage
\begin{figure}
\epsscale{1.0}
\plotone{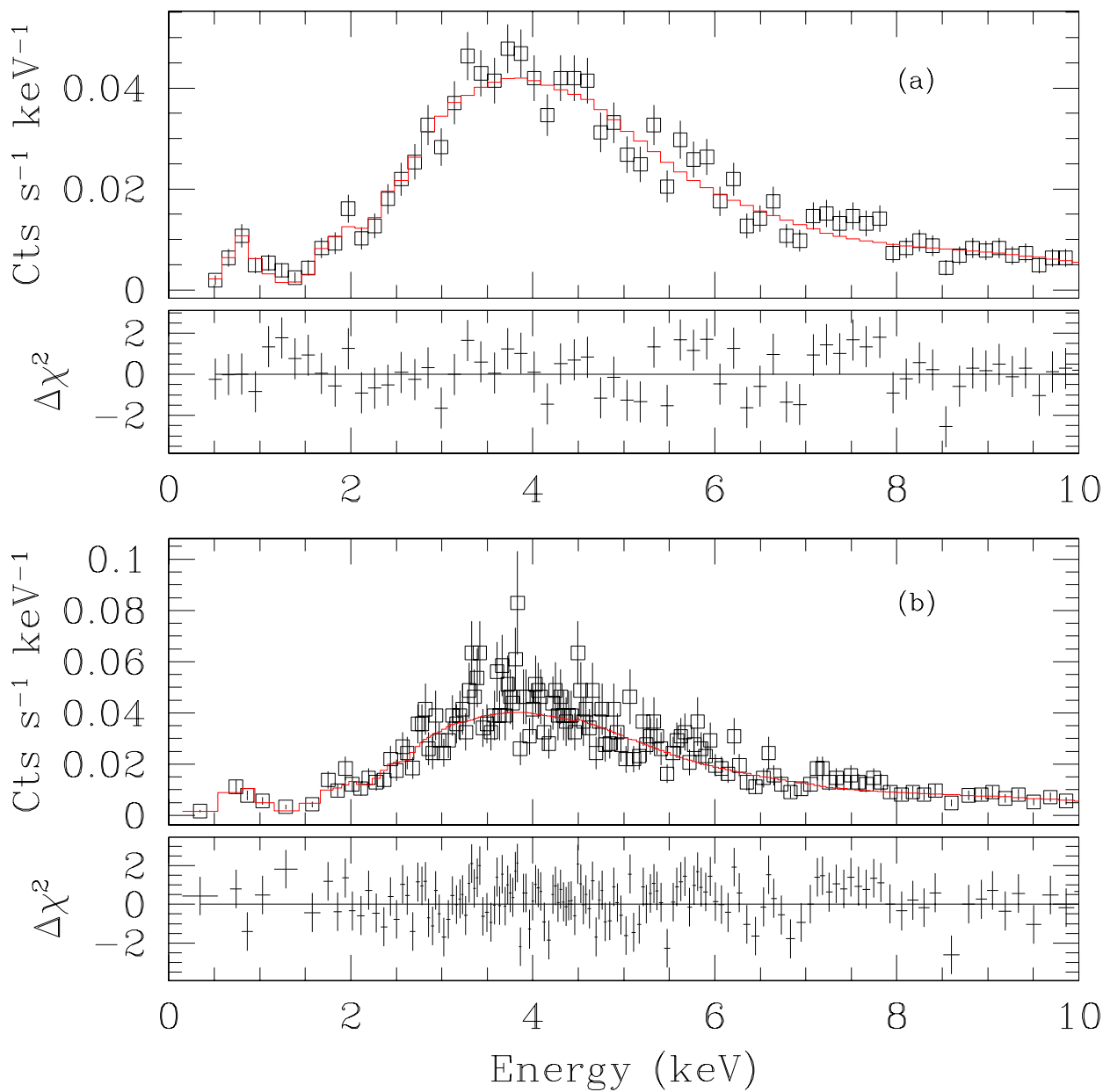}
\caption{Same as Fig.~\ref{f:spectrum1_chandra} for the observation of 2000-04-17. This observation was affected by pile-up, of $\sim 30\%$, which has been included in the modeling. Data were ``grouped'' by 10 pulse-invariant channels (panel $a$) or by a minimum number of counts of 15 per channel (panel $b$). The latter grouping scheme was the same as that used by Young \& Wilson (2004) for the same dataset.}
\label{f:spectrum2_chandra}
\end{figure}

\newpage
\begin{figure}
\epsscale{1.0}
\plotone{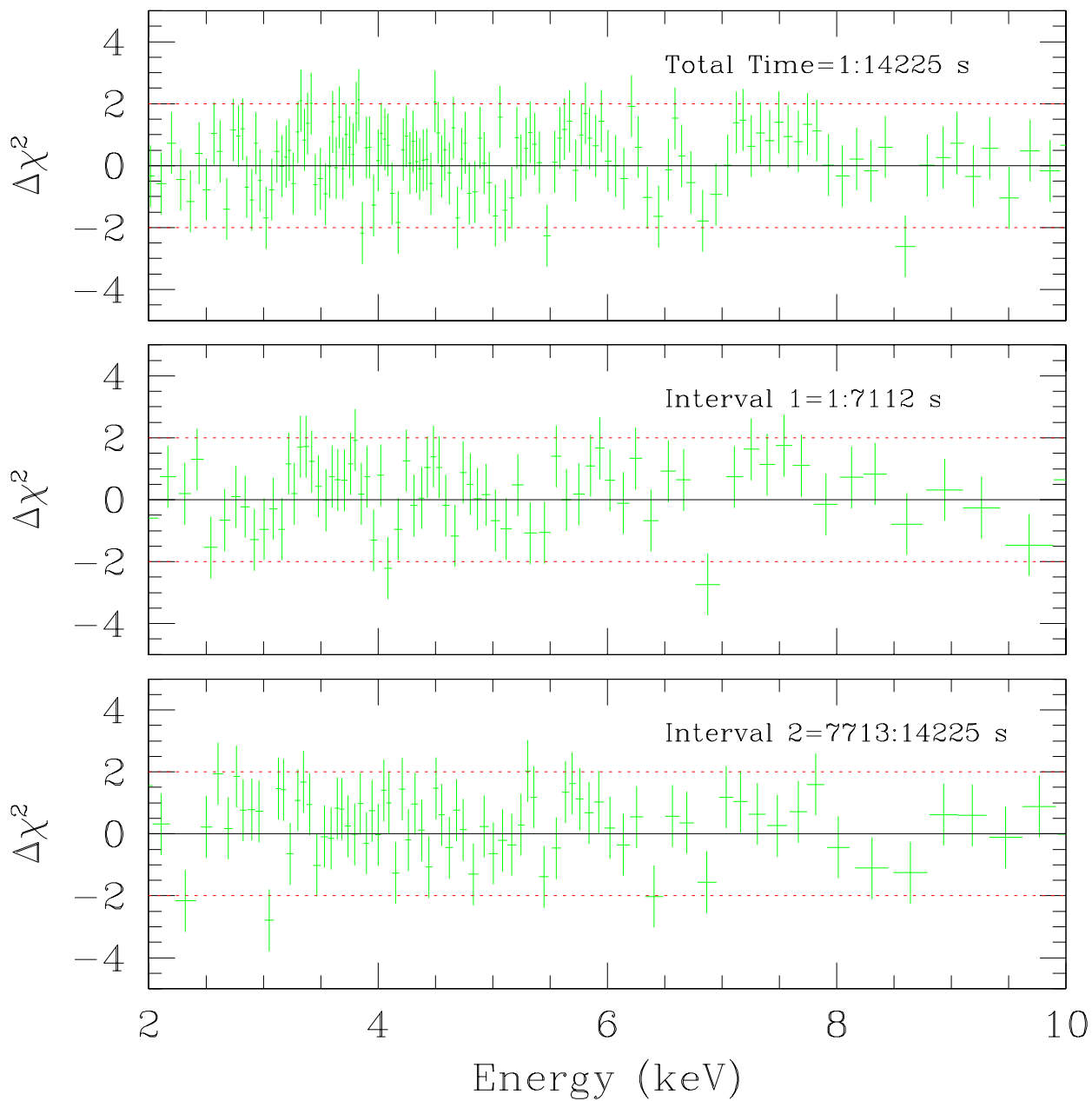}
\caption{Residuals (in units of $\Delta\chi^2 (\sigma)$) to an absorbed power law plus thermal plasma model for the 2000-04-17 observation. The figure compares directly to Figure 5 from Young \& Wilson (2004): data were grouped by a minimum number of counts per channel of 15, and the three panels represent the entire set of observations, and the same divided into two equal $\sim 7000$~s segments. The 2$\sigma$ levels are shown as dotted lines.}
\label{f:res2_chandra}
\end{figure}

\newpage
\begin{figure}
\epsscale{1.0}
\plotone{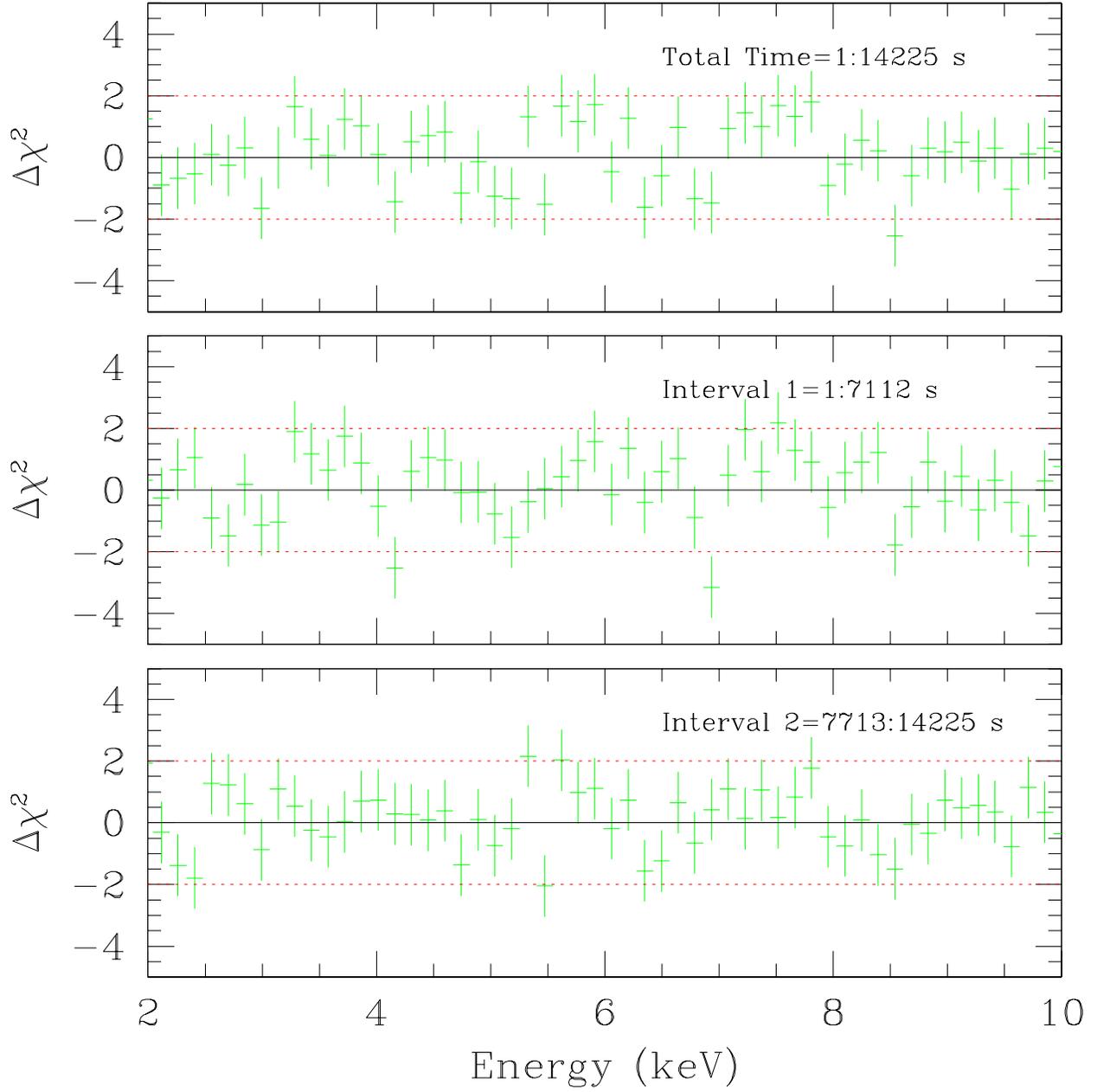}
\caption{Same as Fig.~\ref{f:res2_chandra} with grouping by 10 PI channels.}
\label{f:res2_bin10_chandra}
\end{figure}

\newpage
\begin{figure}
\epsscale{1.0}
\plotone{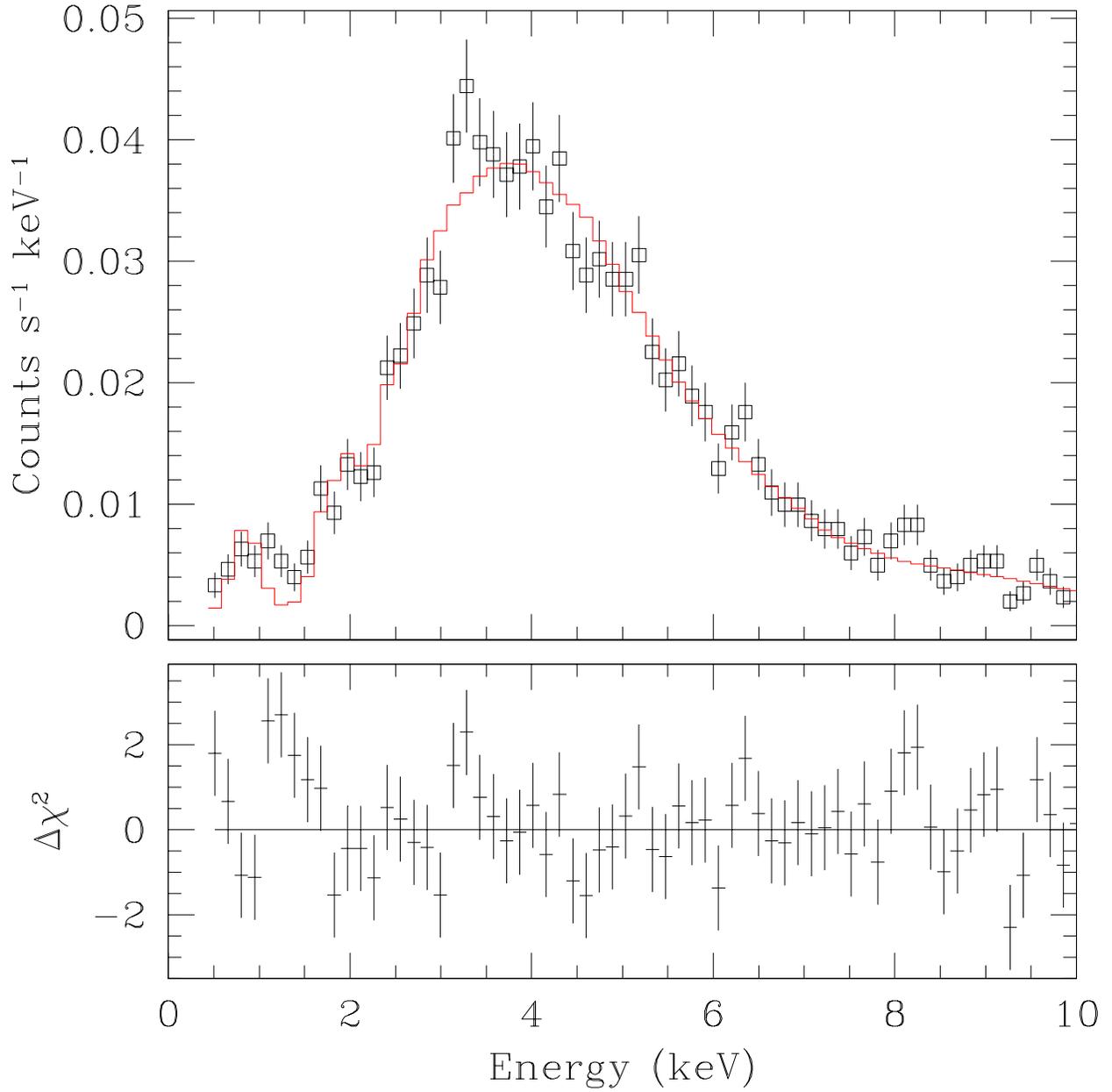}
\caption{Same as Fig.~\ref{f:spectrum1_chandra} for the observation of 2001-05-28. This observation was affected by pile-up 
of $\sim 16\%$.}
\label{f:spectrum3_chandra}
\end{figure}

\newpage
\begin{figure}
\epsscale{1.0}
\plotone{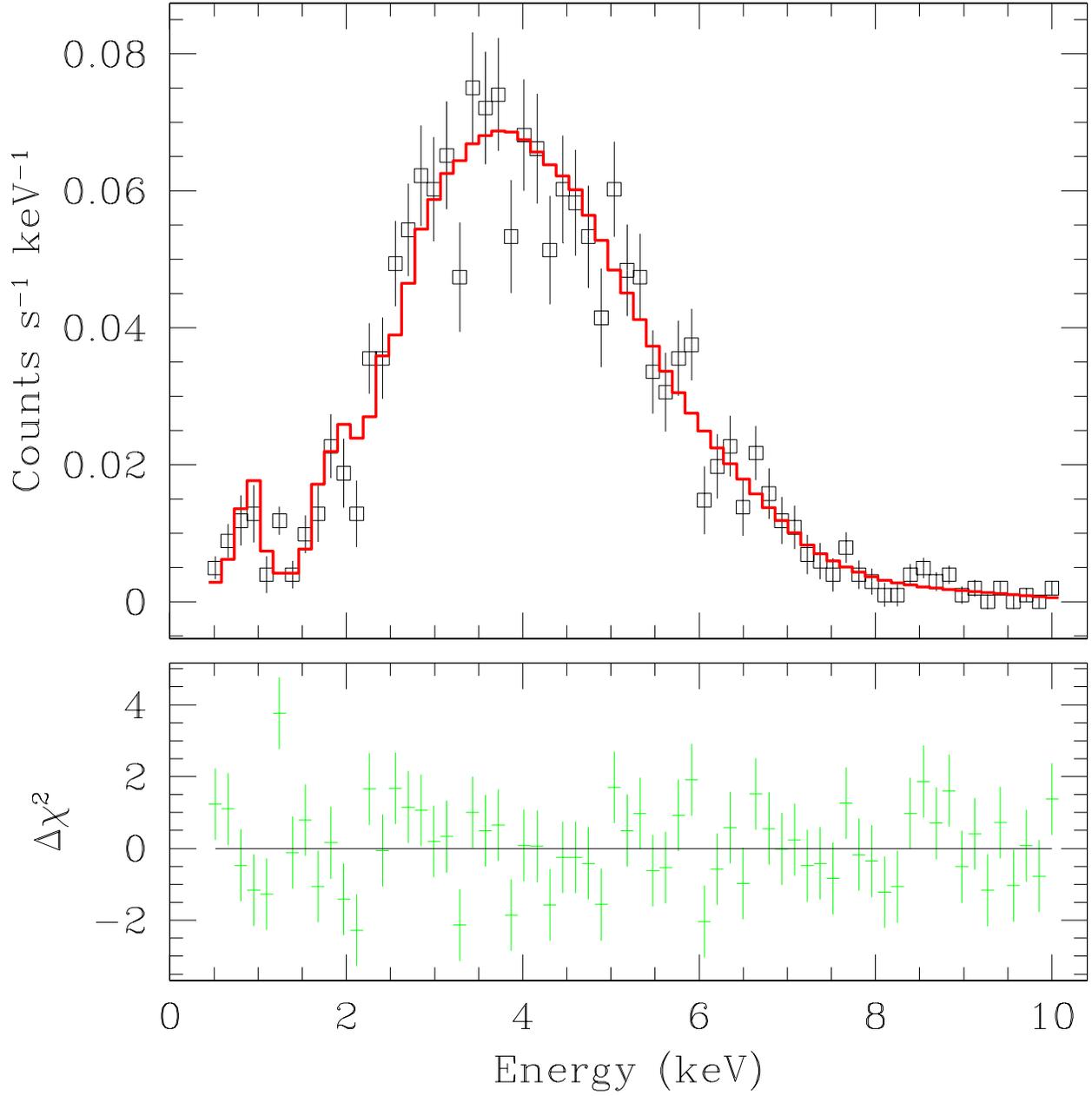}
\caption{Same as Fig.~\ref{f:spectrum1_chandra} for the observation of 2001-05-29.}
\label{f:spectrum4_chandra}
\end{figure}

\newpage
\begin{figure}
\epsscale{1.0}
\plotone{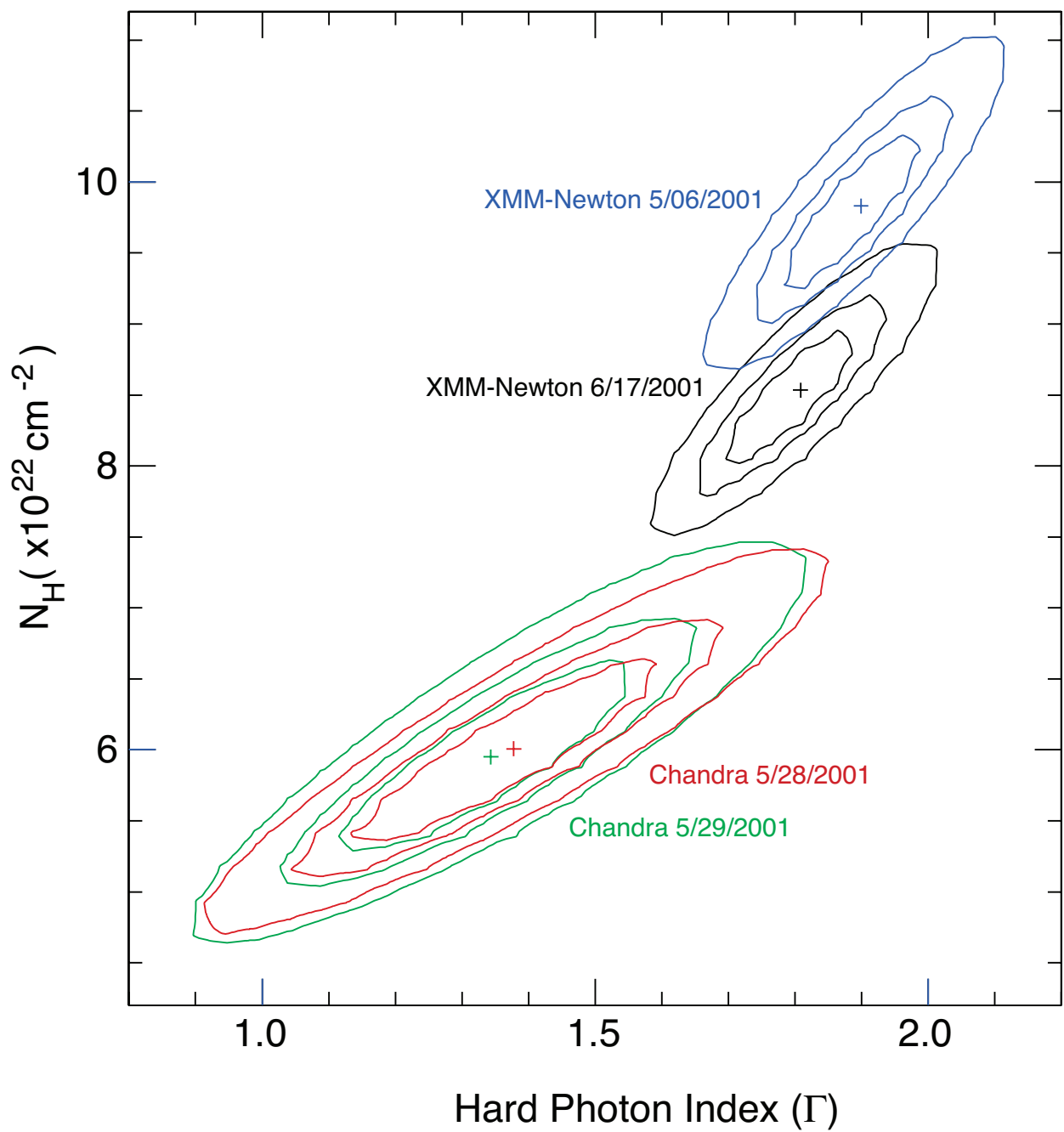}
\caption{Contour plots of confidence regions for the model column density ($N_H$) and photon index ($\Gamma$) parameters. Contours for two sample \XMM\ and \Chandra\ observations are overlaid in the figure.
Confidence intervals for each parameter are estimated independently
and the three level represent confidence of 68.3\%, 90\% and 99\%
(corresponding to 1, 1.6 and 2.6 $\sigma$ respectively). The plus sign marks the best-fit value.}
\label{f:confidence}
\end{figure}

\newpage
\begin{figure}
\epsscale{0.8}
\plotone{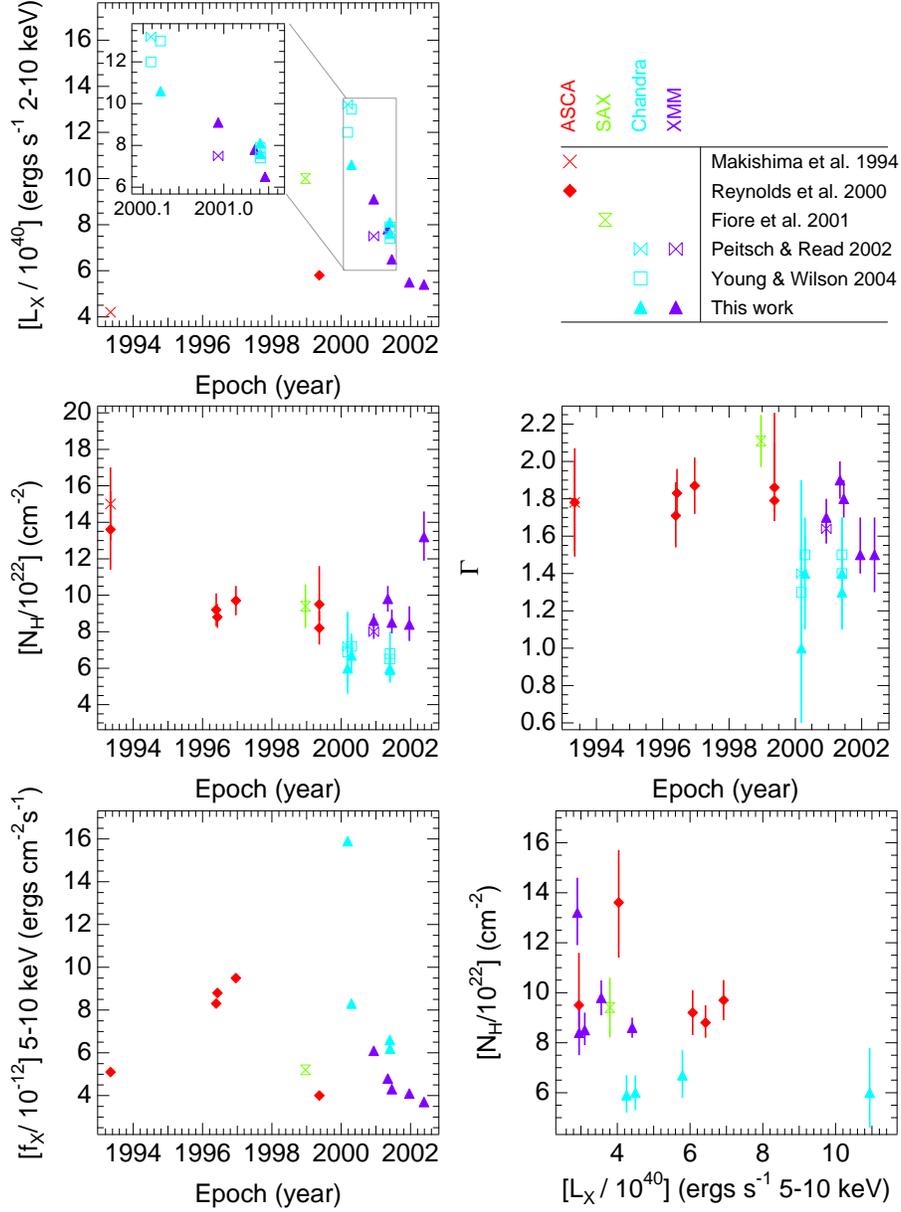}
\caption{
Time series of unabsorbed luminosity ($L_X$), 
model column density ($N_H$), photon index ($\Gamma$),
and absorbed flux ($f_X$), as listed in
Table~2.  The {\it lower right} panel shows $N_H$ vs. hard $L_X$, which
tracks the hard $f_X$ closely.  No correlation is apparent, indicating
that variations in flux are intrinsic to the central engine.  Color is
used to indicate mission while symbol type is used to code the
reporting authors.  Each dataset has been processed by more than one
team.  Plots of
$L_X$(2--10~keV) and $f_X$(5--10~keV) are more sparse than plots of
$N_H$ and $\Gamma$ because no single band is used throughout the
literature.}
\label{f:trends}
\end{figure}

\end{document}